\documentclass[aps,prc,twocolumn,showpacs,amsfonts,amsmath,amssymb,nofootinbib,superscriptaddress]{revtex4-1}
\usepackage{graphicx} 
\usepackage{dcolumn}

\begin{document}

\title{
High-spin states with seniority $v=4,5,6$ in $^{119-126}$Sn 
}

\author{A.~Astier}
\author{M.-G. Porquet}
\affiliation{CSNSM, CNRS/IN2P3 and Universit\'e Paris-Sud, B\^at 104-108,
F-91405 Orsay, France}
\author{Ch.~Theisen}
\affiliation{CEA, Centre de Saclay, 
IRFU/Service de Physique Nucl\'eaire, F-91191 Gif-sur-Yvette Cedex, France}
\author{D.~Verney}
\affiliation{IPNO, CNRS/IN2P3 and Universit\'e Paris-Sud, F-91406 Orsay, France}
\author{I.~Deloncle}
\affiliation{CSNSM, CNRS/IN2P3 and Universit\'e Paris-Sud, B\^at 104-108,
F-91405 Orsay, France} 
\author{M.~Houry}
\altaffiliation{Present address: CEA/DSM/D\'epartement de recherches sur la Fusion
Contr\^ol\'ee, F-130108 Saint-Paul lez Durance, France}
\affiliation{CEA, Centre de Saclay, 
IRFU/Service de Physique Nucl\'eaire, F-91191 Gif-sur-Yvette Cedex, France}
\author{R.~Lucas}
\affiliation{CEA, Centre de Saclay, 
IRFU/Service de Physique Nucl\'eaire, F-91191 Gif-sur-Yvette Cedex, France}
\author{F.~Azaiez}
\altaffiliation{Present address: IPNO,  IN2P3-CNRS and Universit\'e Paris-Sud, 
F-91406 Orsay, France}
\affiliation{IPHC, IN2P3-CNRS and Universit\'e Louis Pasteur, F-67037 Strasbourg 
Cedex 2, France}
\author{G.~Barreau}
\affiliation{CENBG, IN2P3-CNRS and Universit\'e Bordeaux I, F-33175 Gradignan, France}
\author{D.~Curien}
\affiliation{IPHC, IN2P3-CNRS and Universit\'e Louis Pasteur, F-67037 Strasbourg Cedex 2, France}
\author{O.~Dorvaux}
\author{G.~Duch\^ene}
\affiliation{IPHC, IN2P3-CNRS and Universit\'e Louis Pasteur, F-67037 Strasbourg Cedex 2, France} 
\author{B.J.P.~Gall}
\affiliation{IPHC, IN2P3-CNRS and Universit\'e Louis Pasteur, F-67037 Strasbourg Cedex 2, France}
\author{N.~Redon}
\affiliation{IPNL, IN2P3-CNRS and Universit\'e Claude Bernard, F-69622 Villeurbanne Cedex, France} 
\author{M.~Rousseau}
\affiliation{IPHC, IN2P3-CNRS and Universit\'e Louis Pasteur, F-67037 Strasbourg Cedex 2, France}
\author{O.~St\'ezowski}
\affiliation{IPNL, IN2P3-CNRS and Universit\'e Claude Bernard, F-69622 Villeurbanne Cedex, France}

\date{\hfill \today}

\begin{abstract}
The $^{119-126}$Sn nuclei have been produced as fission fragments 
in two reactions induced by heavy ions: $^{12}$C+$^{238}$U at 90 MeV 
bombarding energy, $^{18}$O+$^{208}$Pb at 85 MeV. 
Their level schemes have been built from gamma rays detected using the 
Euroball array. High-spin states located above the long-lived 
isomeric states of the even- and odd-A $^{120-126}$Sn 
nuclei have been identified. Moreover isomeric states 
lying around 4.5 MeV have been established in $^{120,122,124,126}$Sn from the 
delayed coincidences between the fission fragment detector SAPhIR and 
the Euroball array. The  states located above 3-MeV excitation energy are 
ascribed to several broken pairs of neutrons occupying the 
$\nu h_{11/2}$ orbit. The maximum value of angular momentum available in such a
high-j shell, i.e. for mid-occupation and the breaking of the three neutron
pairs,  has been identified. This process 
is observed for the first time in spherical nuclei.
\end{abstract} 
\maketitle
\section{Introduction}\label{introduction}
Experimental and theoretical investigations of the structure of the 
$_{50}$Sn nuclei  have been the subject of much interest during the last decades. 
Their low-lying states are textbook examples of shell model approaches, as their
description only involves excitations of a few neutrons, all along the chain of 
known isotopes, $^{101-134}$Sn.
  
The high-spin states of $^{A}$Sn nuclei with A $< 120$ can be populated by
fusion-evaporation reactions induced by heavy ions. Such experiments, 
performed many years ago, mainly led to the identification of collective
rotational bands up to spin $\sim 20 \hbar$~\cite{br79}. These bands are 
built on 'intruder'
configurations, i.e. two-particle two-hole excitations across the $Z=50$ closed 
shell. At mid neutron shell, this configuration is low in energy, thus 
the collective rotational band being yrast in the mass range 110-118, dominates
the high-spin level schemes.  
Since this is no longer the case for A $\ge$ 120, the yrast 
states of the heavy Sn isotopes are expected
to be only due to excitations of neutrons moving in a spherical well, particularly
the states due to the breaking of several pairs in the $\nu h_{11/2}$ orbit.
Unfortunately  
because of the lack of suitable stable projectile-target combinations, 
high-spin states of heavy Sn isotopes cannot be populated by 
fusion-evaporation reactions. Thus up to now, medium spin states of the 
$^{120-126}$Sn isotopes were only
measured up to spin  $I^\pi = 10^+$ for the even mass and $I^\pi = 27/2^-$ for
the odd mass, by using reactions induced by light ions, deep inelastic reactions,
isomeric decays of  long-lived states of Sn produced by fission of actinides,
or $\beta$-decays of the high-spin long-lived states of heavy 
In~\cite{da86,ma94,br92,pi00,zh00,lo08,fo81}. During the completion of the present 
work,
the decay of a new isomeric state in $^{128}$Sn populated in the fragmentation of 
$^{136}$Xe has been reported~\cite{pi11}, which has been proposed to be the 
15$^-$ state expected from the $(\nu h_{11/2})^{-3}(\nu d_{3/2})^{-1}$ 
configuration. 

For the studies presented in this paper, the $^{119-126}$Sn isotopes
have been
produced as fragments of binary fission induced by heavy ions. We have selected
two fusion-fission reactions in order to identify unambiguously the $\gamma$-rays emitted
by the high-spin states of these nuclei. Moreover $\gamma- \gamma$ angular
correlations have been analyzed in order to assign spin and parity values to most
of these states.  
In addition, new isomeric states lying
around 4.5~MeV have been established in $^{120,122,124,126}$Sn from the 
delayed coincidences between fission fragment detectors and the gamma array. 
All the observed states can be described in terms of broken neutron pairs  
occupying the $\nu h_{11/2}$ orbit. The maximum value of angular momentum 
available in this high-j shell, i.e. for mid-occupation and the breaking of 
the three pairs, has been identified.

\section{Experimental methods and data analysis}\label{experiment}
\subsection{Reactions and $\gamma$-ray detection}
The $^{12}$C + $^{238}$U reaction was studied at 90 MeV incident energy. The beam was  
provided by  the Legnaro XTU tandem accelerator. The 47 mg/cm$^{2}$ target 
of $^{238}$U was thick enough to stop the recoiling nuclei. 
The second reaction, $^{18}$O + $^{208}$Pb at 85 MeV beam energy, was studied at the 
Vivitron accelerator of IReS (Strasbourg). The thickness of the target was 
100 mg/cm$^{2}$. In these two experiments, 
the gamma-rays were detected with the Euroball  
array consisting of 71 Compton-suppressed Ge detectors~\cite{si97} (15 cluster germanium 
detectors placed in 
the backward hemisphere with respect to the beam, 26 clover germanium detectors located 
around 90$^\circ$, and 30 tapered single-crystal germanium detectors located at forward 
angles). 
Each cluster detector is composed of seven closely packed large-volume Ge crystals 
\cite{eb96} and each 
clover detector consists of four smaller Ge crystals \cite{du99}. 
The data were recorded in an event-by-event mode with the requirement that 
a minimum of 
five (three) unsuppressed  Ge detectors fired in prompt coincidence (within a 
time window of 50~ns) during the first (second)
experiment. About 1.9$\times$10$^9$ (4$\times$10$^9$)
coincidence events (within a time window of 300~ns) with a $\gamma$ multiplicity greater than or equal to three 
were registered. 
The offline analysis consisted of both multi-gated 
spectra and  several three-dimensional "cubes" built 
and analyzed with the Radware package \cite{ra95}.

\subsection{ Isomer selection}
To identify new isomeric states in fission fragments, we have performed another
experiment using a fission fragment 
detector to trigger the Euroball array and isolate the delayed $\gamma$-ray cascades. 
The heavy-ion detector, SAPhIR\footnote{SAPhIR, Saclay Aquitaine Photovoltaic cells
for Isomer Research.}, is made of many photovoltaic cells which can be arranged 
in several geometries \cite{th98}. In the present work, it consisted of 32 photovoltaic 
modules laying in four rings around the target. We have used the $^{12}$C + $^{238}$U 
reaction at 90 MeV with a thin target, 0.14 mg/cm$^{2}$. Fragments escaping from the
target are stopped in the photovoltaic cells of SAPhIR. The detection of the two
fragments in coincidence provides a clean signature of fission events. The Euroball
time window was [50~ns--$1\mu$s], allowing detection of delayed $\gamma$-rays emitted during the
de-excitation of isomeric states. 

Time spectra between 
fragments and $\gamma$-rays were analyzed in order to measure the half-life of 
isomeric levels. The FWHM of the time distribution for prompt $\gamma$-rays was around
15 ns. In this experiment, new isomeric states were 
found in $^{120,122,124}$Sn nuclei, which will be detailed below.
\subsection{Identification of new $\gamma$-ray cascades}

The fusion-fission channel of the above-mentioned reactions leads to the production of the high-spin states
of $\sim$150 fragments, mainly located on the neutron-rich side of the valley of stability.
This gives 
several thousands of $\gamma$ transitions which have to be sorted out. Single-gated 
spectra are useless in the majority of cases. The selection of one particular nucleus 
needs at least two energy conditions, implying that at least two transitions have to 
be known. 

The identification of transitions depopulating high-spin levels which are completely 
unknown is based on the fact that {\it prompt} $\gamma$-rays emitted by complementary 
fragments are detected in coincidence \cite{ho91,po96}. For each reaction used in this 
work, we have studied  the intensities of $\gamma$-rays emitted by many pairs 
of complementary fragments with  known  
cascades to establish the relationship between  their number of protons and neutrons. 
The sum of the proton numbers of complementary fragments has been found to be  
the atomic number of the compound 
nucleus\footnote{In the $^{12}$C + $^{238}_{92}$U reaction, we have also identified a weak exit
channel: Few pairs 
of fragments having $Z_1+Z_2=94$, instead of 98, indicates a fission process occuring 
after the transfer of a few nucleons.}, so that the $_{50}$Sn isotopes are 
associated to the $_{48}$Cd isotopes in the $^{12}$C + $^{238}$U reaction and to the
$_{40}$Zr isotopes in the $^{18}$O + $^{208}$Pb reaction. 
The number of evaporated neutrons 
(sum of the pre- and post-fission emitted neutrons) extends from 7 to 14  
in the first reaction \cite{ho99,ho00} and from 2 to 7 in the 
second one \cite{lu02,po04a}. The distribution and the mean number of emitted
neutrons depend slightly on the $N/Z$ ratio of the investigated fragments and 
on the angular momentum of their excited states emitting the $\gamma$-rays. 
Primary fragments populated at
high excitation energy cool down predominantly by neutron evaporation, then
the secondary fragments emit $\gamma$-rays. Therefore the number of emitted
neutron must
be low in order to observe $\gamma$-ray cascades in the most neutron-rich 
isotopes. 

Many new $\gamma$-ray cascades of the $^{120-126}$Sn nuclei
have been identified using the distribution of masses of their partners, as 
explained below.  
It is worth noting that the use of two different reactions to produce the various Sn 
isotopes has turned out to be essential to disentangle the coincidence 
relationships which are often complicated by the existence of many doublets or 
triplets of transitions very close in energy. 

The relative intensity of the lowest transitions in the new cascades identified
in $^{120-126}$Sn have been
measured in the spectra in double coincidences with one new transition and one
transition of a partner. As for the other transitions, we have used spectra in 
double coincidences with two transitions of the new cascades. 
A loss in intensity occurs when going through an isomeric state. 
Knowing that the time window was 300~ns for the two experiments, 
such an effect has been taken into account for the half-lives in the 100-300~ns range,
observed in the present work.

\subsection{$\gamma$-$\gamma$ angular correlations \label{correl}}
In order to determine the spin values of excited states,
the coincidence rates of two successive $\gamma$ transitions are 
analyzed as a function of $\theta$, the
average relative angle between the two fired detectors.
The Euroball spectrometer had $C^{2}_{239}$~=~28441 
combinations of 2 crystals, out of which only $\sim$ 2000  
involved different values of relative angle within 2$^\circ$. 
Therefore, in order 
to keep reasonable numbers of counts, all the angles have been 
gathered around three average relative angles : 22$^\circ$, 46$^\circ$, 
and 75$^\circ$. 

The coincidence rate is increasing between 0$^\circ$ and 
90$^\circ$ for the dipole-quadrupole cascades, whereas it decreases for 
the quadrupole-quadrupole 
or dipole-dipole ones. More precisely, the angular correlation functions 
at the three angles 
of interest were calculated for several combinations of spin sequences,
corresponding to typical multipole orders (see table \ref{correl_th}). 
In order to check the method, angular correlations of transitions 
belonging to the yrast 
cascades of the fission fragments having well-known multipole orders 
were analyzed and the expected values were found in all cases. 
\begin{table}[!h]
\begin{center}
\caption{Values of the angular correlation functions, R($\theta$), normalized 
to the ones calculated at 75$^\circ$, computed for several combinations of 
spin sequences and multipole orders (Q = quadrupole, D = Dipole).
}
\label{correl_th}
\begin{ruledtabular}
\begin{tabular}{ccccc}
Spin sequence & Multipole  &R(22$^\circ$)&R(46$^\circ$)&R(75$^\circ$)\\
$I_1-I_2-I_3$ & orders& & & \\
\hline
14 -- 12 -- 10& Q - Q  &  1.13& 1.06	&1.00	\\
12 -- 11 -- 10& D - D &  1.06& 1.03	&1.00	\\
13 -- 12 -- 10& D - Q &  0.92& 0.96	&1.00	\\
\end{tabular}
\end{ruledtabular}
\end{center}
\end{table}

When the statistics of our data are too low to perform
such a measurement, the spin assignments are based upon 
(i) the already known spins of some states, (ii) the assumption 
that in yrast decays, spin values increase with the excitation energy, 
(iii) the possible existence of cross-over transitions, and
(iv) the analogy with the level structures of the other isotopes. 
 
\section{Experimental results}\label{results}

The $\gamma$-rays emitted by the low-lying states of $^{118-126}$Sn isotopes
have been observed in both fusion-fission reactions used in the present work.
As for $^{118}$Sn, its yield is so low as only the decay of its 7$^-$ state 
is observed and the transitions of its partners could not be identified. 
On the other hand, we have measured many new $\gamma$-rays
emitted by the high-spin states of $^{119-126}$Sn, the results are
presented in the three following sections.  

\subsection{New $\gamma$-ray cascades of $^{120-126}$Sn isotopes}\label{identification}

Up to now, the high-spin level schemes of $^{120-126}$Sn isotopes could only 
be established up to the long-lived isomeric states, with I$^\pi$ = 10$^+$ or 
27/2$^-$ at about 2-3~MeV excitation energy, since 
it is difficult to register coincidence relationships between the transitions populating 
and depopulating states having such half-life values (T$_{1/2}=1-62~\mu s$) using
standard experimental apparatus. On the other hand, in the fusion-fission experiments, 
all the $\gamma$-ray cascades located above the long-lived isomeric states of the Sn isotopes 
are easily detected in prompt coincidences with those emitted by their complementary 
fragments. Three steps are necessary to carry out the search: 
\begin{itemize}
\item to find all the $\gamma$ rays emitted in
prompt coincidences with the transitions of the $^{114-122}$Cd fragments in the  
first data set and the $^{96-100}$Zr fragments in the second one, which 
do not belong to their respective level schemes.
\item to build all the $\gamma$-ray cascades using their own coincidences.
\item to assign each $\gamma$-ray cascade to one particular Sn isotope, from the distribution
of masses of their partners. 
\end{itemize}
 
\begin{figure}[!h]
\includegraphics*[width=8.5cm]{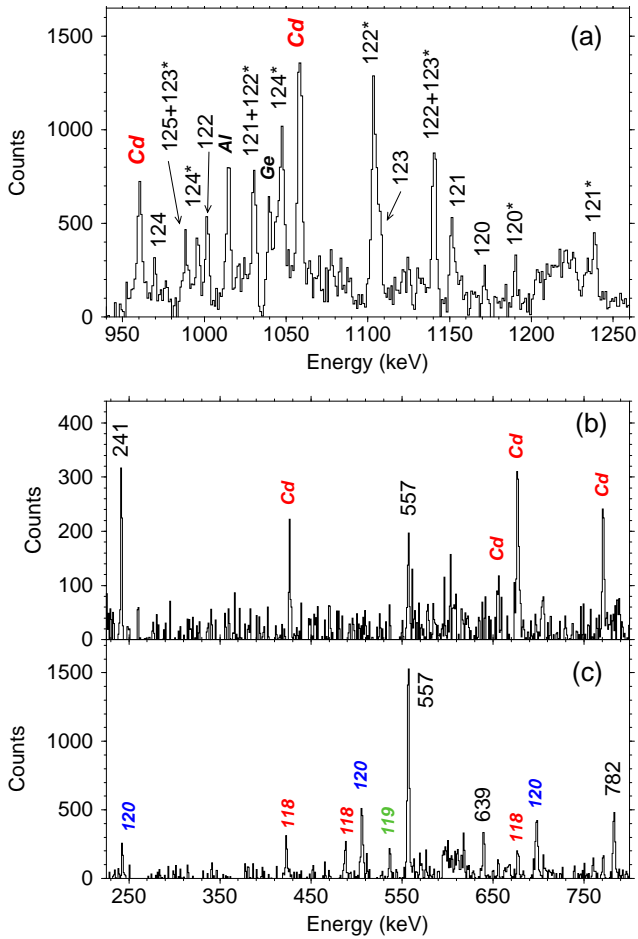}
\caption{(Color online) (a) Spectrum of $\gamma$ rays detected in coincidence 
with the first two
transitions of $^{118}$Cd (487 and 676~keV), built from the data set of the  
$^{12}$C + $^{238}$U reaction, in the [950-1250~keV] energy range. 
All the transitions but two ones, emitted by $^{118}$Cd, belong to 
its complementary fragments, $^{120-125}$Sn.
They are labelled by their mass A, an
asterisk being added when the transition belongs to a new cascade located above
a long-lived isomeric state. 
(b) Spectrum of $\gamma$ rays detected in coincidence 
with the first transition of $^{118}$Cd (487~keV) and the new transition at
1190~keV. Transitions emitted by $^{118}$Cd are labelled by Cd. 
(c) Spectrum of $\gamma$ rays detected in coincidence 
with the two new transitions at 1190~keV and 241~keV. Transitions emitted 
by the $^{118-120}$Cd complementary fragments are labelled by their mass. 
}
\label{spectreCd118}      
\end{figure}

The first item  is illustrated in Fig.~\ref{spectreCd118}(a), gated
by the first two  transitions of  $^{118}$Cd (487 and 676~keV), built from the data set of the  
$^{12}$C + $^{238}$U reaction.
Besides two transitions already known in $^{118}$Cd in the [950-1250-keV] energy 
range, a lot of new transitions are observed and can be assigned to 
$^{120-125}$Sn, half of them being known to belong to the cascades built on 
their ground state (they are labelled by their mass A). Then in order to find the
$\gamma$-ray cascades comprising the other transitions, all the spectra in double
coincidence with one transition of $^{118}$Cd and one new transition are
precisely analyzed. For instance in  
Fig.~\ref{spectreCd118}(b), two new $\gamma$ lines at 241~keV and 557~keV are found 
to be correlated to the new 1190-keV transition and the 487~keV $\gamma$-ray 
of $^{118}$Cd. Finally,  
Fig.~\ref{spectreCd118}(c) allows to complete the search of a new $\gamma$-ray 
cascade: It
comprises five transitions at 1190~keV, 557~keV, 241~keV, 782~keV and 639~keV. 
In addition, the fact that this third spectrum shows the transitions emitted 
by {\bf several} Cd complementary fragments confirms that this cascade belongs to 
the level scheme of one Sn isotope. 
Using the same procedure for each Cd isotope in the first
experiment, as well as for $^{96-100}$Zr in the second one, several new $\gamma$-ray 
cascades have been observed in the present work. 

\begin{figure}[!h]
\includegraphics*[width=7cm]{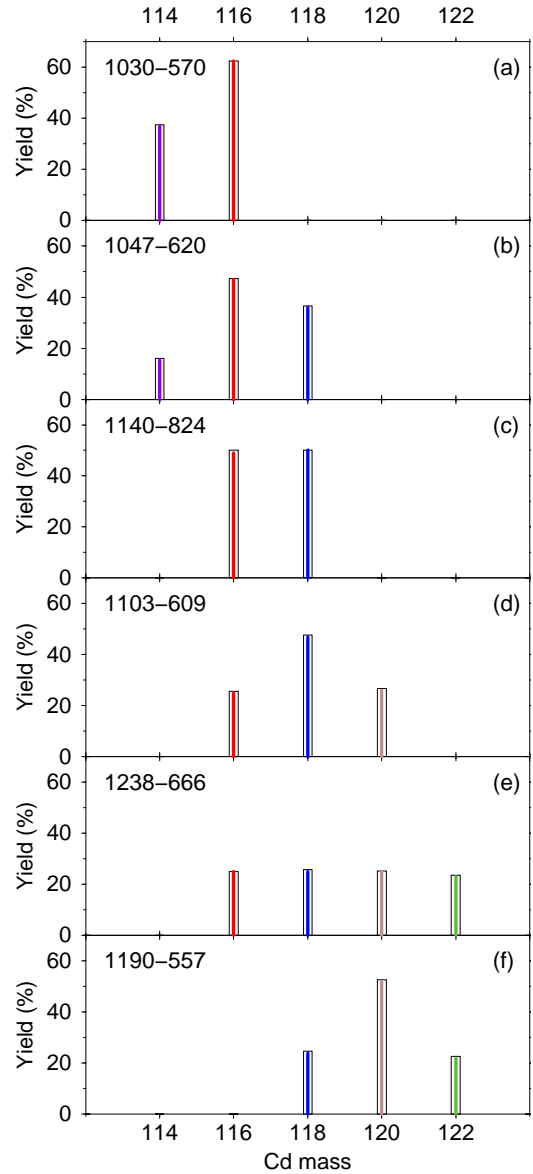}
\caption{(Color online) Relative yields of even-$A$ Cd isotopes associated to the new 
$\gamma$-ray cascades emitted by the Sn isotopes produced in the 
$^{12}$C + $^{238}$U reaction. The yield of each even-$A$ Cd isotope is computed 
from the number of counts of its $2^+ \rightarrow 0^+$ $\gamma$ line in the 
spectra gated by the two strongest transitions of the new cascades (their energies 
are written in each drawing). 
(a) case of the new cascade assigned to $^{126}$Sn, (b) $^{124}$Sn, (c) $^{123}$Sn,
(d) $^{122}$Sn, (e) $^{121}$Sn, (f) $^{120}$Sn.
The new cascades assigned to $^{119}$Sn and $^{125}$Sn are discussed in
Sec.~\ref{119et125}.}
\label{correl_mass}      
\end{figure}

To assign every $\gamma$-ray cascade to one particular Sn isotope, we have 
first analyzed the relative intensities of $\gamma$-rays emitted by the partners in
each spectrum in double coincidence with two transitions of the new cascades (such as
the spectrum shown in Fig.~\ref{spectreCd118}(c)). Then we have  
analyzed all these distributions of masses of the partners. Examples
of results obtained in the first experiment are shown in Fig.~\ref{correl_mass}.  
The evolution of the Cd yields from the top drawing to the bottom one proves
that the six cascades belong to different Sn isotopes. Similar results are obtained
from the evolution of the Zr yields associated to each cascade, in the second 
experiment. Finally these relative distributions of Cd/Zr masses 
were compared to well known pairs of complementary fragments and the six new
cascades discussed in Fig.~\ref{correl_mass} were assigned to 
$^{120,121,122,123,124,126}$Sn. Their first transitions are
marked with an asterisk in the spectrum of Fig.~\ref{spectreCd118}(a). Their
precise location in the Sn level schemes are discussed in the next sections.

\subsection{Study of the even-$A$ $^{120-126}$Sn isotopes}\label{massepaire}
Very few medium-spin levels were known in the even-$A$ $^{120-126}$Sn isotopes
prior to this work. Populated in deep inelastic reactions~\cite{br92,zh00}, 
two long-lived isomeric 
states were identified from their $\gamma$-decays to the low-lying states. 
Lying between 2 and 3~MeV excitation energy, these isomeric states are due to 
the breaking of one neutron pair, the state with $I^\pi = 10^+$ being 
attributed to a 
$(\nu h_{11/2})^2$ configuration and the one with $I^\pi = 7^-$ to a 
$(\nu h_{11/2})(\nu d_{3/2})$ configuration. 
Using the data of the two fusion-fission reactions of the present work, we have
identified the yrast structures of $^{120-126}$Sn located
above their long-lived isomeric states. In the following, we first present 
the building of each high-spin level scheme  and the measurement of isomeric 
states in the [30-300~ns] range. Then, we discuss the angular momentum and 
parity assignments of most of the new states of $^{120-126}$Sn.   

\subsubsection{$^{120}$Sn}
A cascade starting with the 1190-keV and 557-keV transitions is assigned to 
$^{120}$Sn, because of the mass distribution of its complementary fragments 
in the two fusion reactions (Sec.~\ref{identification}). Besides the three
other transitions shown in Fig.~\ref{spectreCd118}(b) and (c), namely the 
241-keV, 782-keV and 639-keV $\gamma$-rays, a few other $\gamma$-lines have been 
found to belong to the cascade, thanks to their coincidence relationships.
The whole set forms two linked branches which are put on the top of 
the  10$^+$ and 7$^-$ states respectively, since two parallel decay paths
located just below the 241-keV transition, 
557+1190 on the one hand and 253+661+1253 on the other hand,
exactly fits the difference in energy between the 10$^+$ and 7$^-$  
\begin{figure}[!h]
\includegraphics*[height=13cm]{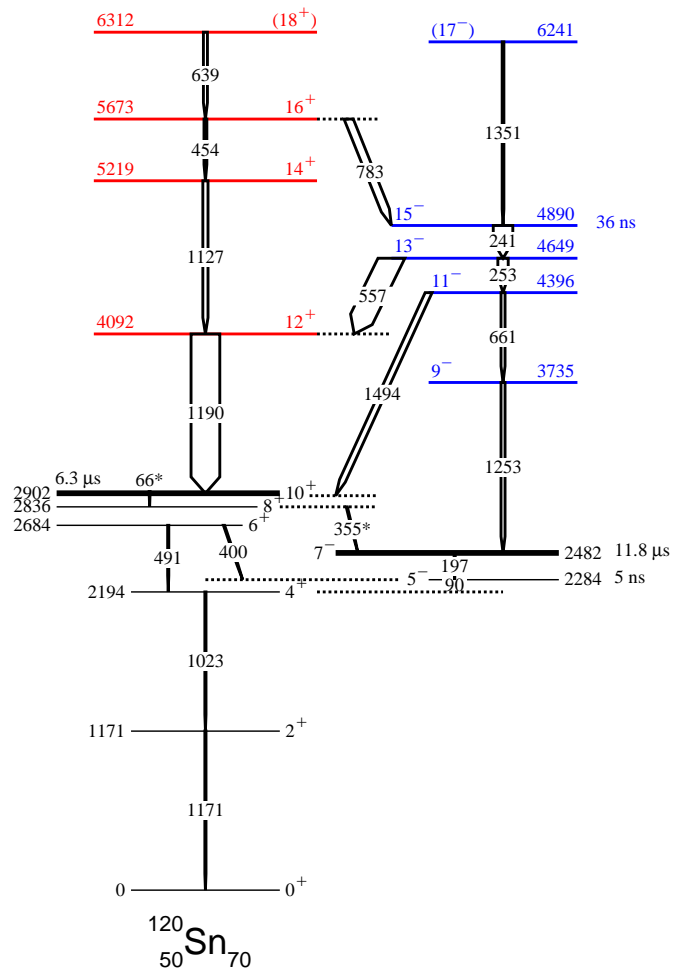}
\caption{(Color online) Level scheme of $^{120}$Sn deduced in the present work. The colored levels are new. The two long-lived isomeric states 
($T_{1/2}=6.26(11) \mu$s and 11.8(5) $\mu$s) and their $\gamma$-decays to the low-lying states were already
known~\cite{NNDC}. The 355-keV transition, located between 
two long-lived isomeric states, as well as the very converted 66-keV transition, 
could not be observed in our work. The width of
the arrows is representative of the relative intensity of the $\gamma$ rays
above the isomeric states. 
}
\label{schema120}      
\end{figure}
\begin{table}[!h]
\caption{Properties of the new transitions assigned to $^{120}$Sn 
in this experiment. The energies of the two long-lived isomeric states
at 2481.6 keV (I$^\pi = 7^-$) and 2902.2 keV (I$^\pi = 10^+$) 
(written in bold) are from 
Ref.~\cite{NNDC}.}
\label{gammas120Sn}
\begin{ruledtabular}
\begin{tabular}{rrccc}
E$_\gamma$(keV)\footnotemark[1]&I$_\gamma$\footnotemark[1]$^,$\footnotemark[2]&J$_i^\pi$$\rightarrow$J$_f^\pi$  &E$_i$(keV)&E$_f$(keV)\\
\hline
241.1(2)  &95(14) &  15$^-$ $\rightarrow$ 13$^-$& 4890.1  & 4649.0 \\
252.9(2)  &36(7) & 13$^-$  $\rightarrow$ 11$^-$	& 4649.0& 4396.0\\
453.7(3)  &9(3)  & 16$^+$  $\rightarrow$ 14$^+$	& 5672.9& 5219.3\\
556.5(2)  &83(12) & 13$^-$  $\rightarrow$ 12$^+$& 4649.0& 4092.5\\
639.0(3)  &14(4) & (18$^+$)$\rightarrow$ 16$^+$	& 6311.9& 5672.9\\				
661.1(3)  &14(4) & 11$^-$  $\rightarrow$ 9$^-$	& 4396.0& 3734.6\\
782.8(3)  &29(6) &  16$^+$ $\rightarrow$ 15$^-$	& 5672.9& 4890.1\\
1127.0(4) &17(4) &  14$^+$ $\rightarrow$ 12$^+$	& 5219.3& 4092.5\\
1190.3(3) &100 &  12$^+$ $\rightarrow$ 10$^+$   & 4092.5& {\bf 2902.2}\\
1253.0(4) &14(4) & 9$^-$   $\rightarrow$ 7$^-$ 	& 3734.6& {\bf 2481.6}\\
1350.8(5) &5(2)  & (17$^-$)$\rightarrow$ 15$^-$ & 6240.9& 4890.1\\
1493.8(4) &22(5) & 11$^-$  $\rightarrow$ 10$^+$ & 4396.0& {\bf 2902.2}\\
\end{tabular}
\end{ruledtabular}
\footnotetext[1]{The number in parenthesis is the error in the last digit.}
\footnotetext[2]{The relative intensities are normalized to the sum 
$I_\gamma(557) + I_\gamma(1127)=100$.}
\end{table}
isomeric states of $^{120}$Sn (see Fig.~\ref{schema120}).
All the transitions newly observed  in $^{120}$Sn are given in 
Table~\ref{gammas120Sn}.
The spin and parity of the new states will be discussed and
assigned in Sec.~\ref{spin_assignment}.

\begin{figure}[!h]
\includegraphics*[width=7.5cm]{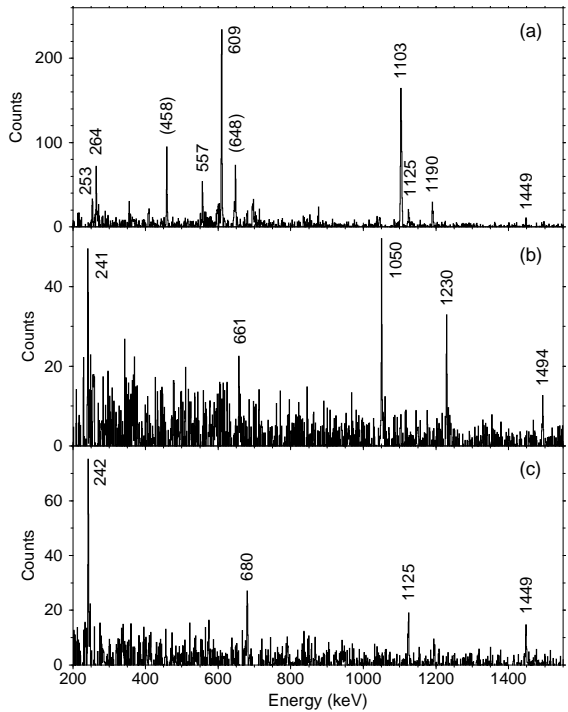}
\caption{Spectra of $\gamma$-rays which have been detected in the
time interval 50~ns-$1 \mu$s after the detection of two fragments
by SAPhIR. (a) $\gamma$-rays in prompt coincidence with the
241/242~keV transition ($^{120}$Sn and $^{122}$Sn). The 458- and 648-keV
transitions are pollutions (they belong to the decay of the isomeric 7$^+$ 
state of $^{88}$Rb~\cite{po09}).
(b) $\gamma$-rays in prompt coincidence with the
253~keV transition, which is a doublet. The 1050- and 1230-keV $\gamma$-rays are the first two
transitions of $^{118}$Sn and the 241-, 661-, and 1494-keV ones are
assigned to $^{120}$Sn.
(c) $\gamma$-rays in prompt coincidence with the 264~keV transition ($^{122}$Sn). 
}
\label{decaysaphir}      
\end{figure}

Using the data from the SAPhIR experiment, the transitions involved in the
de-excitation of the 4890-keV level have been found to be delayed. The spectrum
of $\gamma$-rays which have been detected in the
time interval 50~ns-$1 \mu$s after the detection of two fragments
by SAPhIR and in prompt coincidence with the
241/242~keV transition is drawn in Fig.~\ref{decaysaphir}(a). 
As this $\gamma$ line is a triplet, the spectrum exhibits transitions emitted 
by the isomeric states of three fission fragments, 
$^{88}$Rb~\cite{po09}, $^{120}$Sn and $^{122}$Sn. 
The spectrum shown in Fig.~\ref{decaysaphir}(b) is gated by the 253~keV 
$\gamma$-ray. This line is a doublet, as this energy occurs both in the decay of
the 7$^-$ isomeric state of $^{118}$Sn (associated with the 1050- and 1230-keV
transitions) and in the decay of the isomeric state at 4890 keV, newly 
established in $^{120}$Sn. The statistics of this spectrum is too low to show the
1253-keV transition, identified in the thick-target Euroball experiment, from
double-gated spectra having higher statistics.

\begin{figure}[!h]
\includegraphics*[width=7cm]{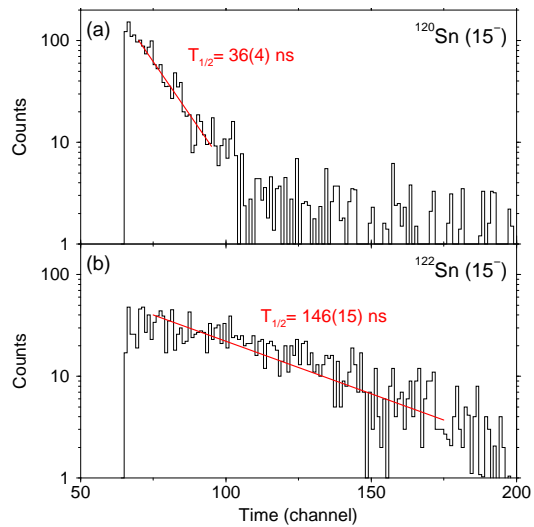}
\caption{(Color online) (a) Half-life of the 4890~keV state of $^{120}$Sn obtained from
the sum of the time distributions of the 1190- and 557-keV transitions.
(b) Half-life of the 4720~keV state of  $^{122}$Sn obtained from
the time distribution of the 1103-keV transition. See text for further details about
the gating conditions and procedures.
}
\label{demivies}      
\end{figure}

The time distribution between the detection of two fragments by SAPhIR and the
emission of the 1190~keV or the 557~keV $\gamma$-ray is shown in
Fig.~\ref{demivies}(a). In order to reduce the background, we have selected
the events containing an additional $\gamma$-ray belonging to the 
241-557-1190 cascade. Several least-squares fits of this spectrum were
performed, varying the time interval from 70 to 100~ns and shifting the smallest
intervals along the time axis. The average of all obtained values is 
$T_{1/2}$= 36(4)~ns, the adopted uncertainty being the observed dispersion 
(greater than the uncertainties quoted for each fit).

\subsubsection{$^{122}$Sn}
\begin{figure}[!h]
\includegraphics*[height=13.5cm]{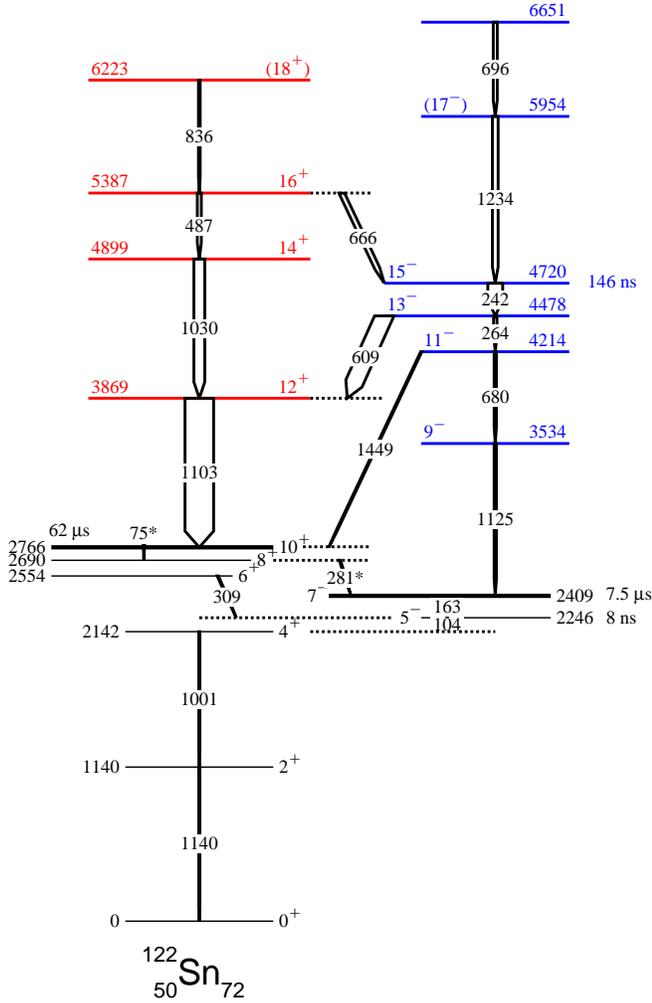}
\caption{(Color online) Level scheme of $^{122}$Sn deduced in the present work. The colored levels
are new.
The two long-lived isomeric states ($T_{1/2}=7.5(9) \mu$s 
and 62(3) $\mu$s) and their $\gamma$-decays to the low-lying states were already
known~\cite{NNDC}. The 281~keV transition, located between 
two long-lived isomeric states, as well as the very converted 75~keV 
transition could not be observed in our work. The width of
the arrows is representative of the relative intensity of the $\gamma$ rays
above the isomeric states.  
}
\label{schema122}      
\end{figure}

As for $^{120}$Sn, the starting point is the two transitions at 1103~keV and
609~keV which have to be placed above its long-lived isomeric states, because
of the mass distribution of its partners 
in the two fusion reactions (Sec.~\ref{identification}). 
Several new transitions are observed in coincidence with the 1103~keV and
609~keV $\gamma$-rays. All the observed relationships result in two linked 
branches which are placed above the the 10$^+$ and 7$^-$
isomeric states of $^{122}$Sn, since two parallel decay paths
located just below the 242~keV transition, 
609+1103 on the one hand and 264+680+1125 on the other hand, 
exactly fits their difference in energy (see Fig.~\ref{schema122}). 
Examples of coincidence spectra of $\gamma$
lines belonging to $^{122}$Sn are given in Fig.~\ref{decaysaphir}(a) and (c),
which demonstrate that all the transitions involved in the
de-excitation of the 4720-keV level are delayed. The time distribution 
between the detection of two fragments by SAPhIR and the
emission of the 1103-keV $\gamma$-ray gives $T_{1/2}$= 146(15)~ns. In order to reduce
the background we have selected the events containing either the 609-keV transition or
the 242-keV one (see Fig.~\ref{demivies}(b)). 

All the transitions newly observed  in $^{122}$Sn are given in 
Table~\ref{gammas122Sn}.
The spin and parity of the new states will be discussed and
assigned in Sec.~\ref{spin_assignment}.
\begin{table}[!h]
\caption{Properties of the new transitions assigned to $^{122}$Sn 
in this experiment. The energies of the two long-lived isomeric states
at 2409.0 keV (I$^\pi = 7^-$) and 2765.6 keV (I$^\pi = 10^+$) (written in bold) 
are from Ref.~\cite{NNDC}.}
\label{gammas122Sn}
\begin{ruledtabular}
\begin{tabular}{rrccc}
E$_\gamma$(keV)\footnotemark[1]&I$_\gamma$\footnotemark[1]$^,$\footnotemark[2]&J$_i^\pi$$\rightarrow$J$_f^\pi$  &E$_i$(keV)&E$_f$(keV)\\
\hline
242.1(2) &52(9) &  15$^-$ $\rightarrow$ 13$^-$   & 4720.5& 4478.4 \\
263.8(4) &11(3) & 13$^-$  $\rightarrow$ 11$^-$	& 4478.4& 4214.4  \\
487.3(3) &14(4) & 16$^+$  $\rightarrow$ 14$^+$	& 5386.7& 4899.4\\
609.3(2) &62(12) & 13$^-$  $\rightarrow$ 12$^+$   & 4478.4& 3869.1\\
665.9(3) &17(4) & 16$^+$ $\rightarrow$ 15$^-$	& 5386.4& 4720.5\\				
680.3(4) &7(3)  & 11$^-$  $\rightarrow$ 9$^-$	& 4214.4& 3534.0 \\
696.3(3) &14(4) &  ~~~~ $\rightarrow$ (17$^-$)	& 6650.8& 5954.5\\
835.9(4) &6(3)  & (18$^+$) $\rightarrow$ 16$^+$  & 6222.6& 5386.7\\
1030.3(3)&38(7) &  14$^+$ $\rightarrow$ 12$^+$	& 4899.4& 3869.1\\
1103.5(3)&100 &  12$^+$ $\rightarrow$ 10$^+$   & 3869.1& {\bf 2765.6}\\
1125.0(4)&7(3)  & 9$^-$   $\rightarrow$ 7$^-$    & 3534.0& {\bf 2409.0}\\
1234.0(4)&21(5) & (17$^-$)$\rightarrow$ 15$^-$   & 5954.5& 4720.5\\
1448.9(4)&4(2)  & 11$^-$  $\rightarrow$ 10$^+$   & 4214.4& {\bf 2765.6}\\
\end{tabular}
\end{ruledtabular}
\footnotetext[1]{The number in parenthesis is the error in the last digit.}
\footnotetext[2]{The relative intensities are normalized to the sum 
$I_\gamma(609) + I_\gamma(1030)=100$.}
\end{table}

\subsubsection{$^{124}$Sn}
\begin{figure}[!h]
\includegraphics*[height=14.3cm]{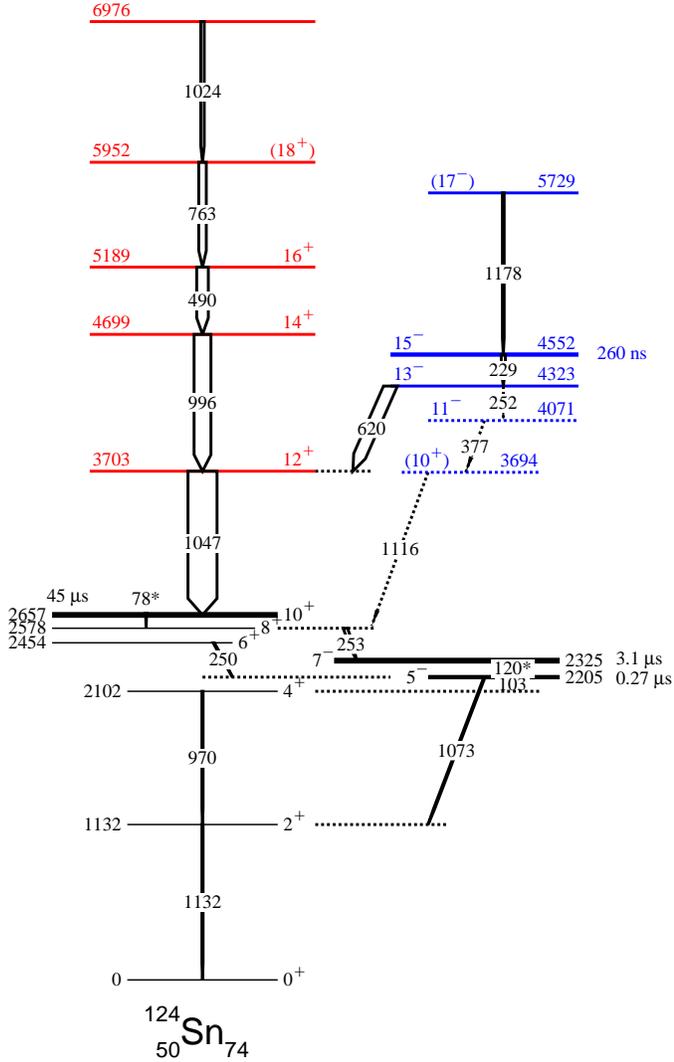}
\caption{(Color online) Level scheme of $^{124}$Sn deduced in the present work. The colored levels
are new. 
The two long-lived isomeric states ($T_{1/2}=3.1(5) \mu$s 
and 45(5) $\mu$s) 
and their $\gamma$-decays to the low-lying states were already
known~\cite{NNDC}. The very converted 78~keV and 120~keV transitions
could not be observed in our work. The width of
the arrows is representative of the relative intensity of the $\gamma$ rays
above the isomeric states.  
}
\label{schema124}      
\end{figure}

Similar procedures were used to identify the high-spin structures of
$^{124}$Sn. By gating on the two coincident transitions at 1047~keV and 620~keV assigned 
to $^{124}$Sn because of the mass distribution of its partners 
in the two fusion reactions (Sec.~\ref{identification}), we found the 229-
and the 1178-keV $\gamma$-rays, establishing a cascade of 4 transitions. 
Using the data from the SAPhIR experiment, the first three transitions 
have been found to be delayed. 
The time distribution between the detection of two fragments by SAPhIR and the
emission of the 1047- or the 620-keV $\gamma$-ray gives $T_{1/2}$= 260(25)~ns.
In order to reduce the background, we have selected
the events containing an additional $\gamma$-ray of the 229-620-1047 cascade.
Since the 229-620-1047 sequence resembles well the
242-609-1103 sequence of $^{122}$Sn and the 241-557-1190 one of $^{120}$Sn,
we have placed the cascade of $^{124}$Sn directly above its 10$^+$ state
at 2657~keV excitation energy (see Fig.~\ref{schema124}).

\begin{figure}[!h]
\includegraphics*[width=8cm]{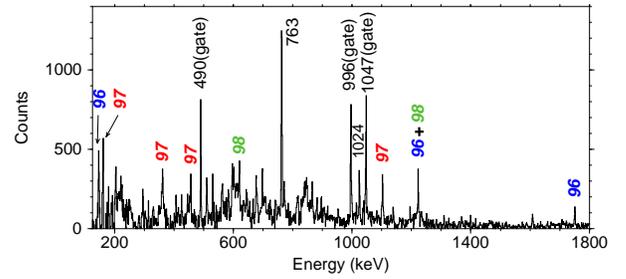}
\caption{(Color online) Coincidence spectrum double-gated on the 1047-, 996-, and
490-keV transitions of a new cascade identified in $^{124}$Sn,
built 
from the $^{18}$O + $^{208}$Pb data set. The $\gamma$-rays emitted
by the Zr complementary fragments are labelled by their masses A,
written in italics.  
}
\label{yrast124Sn}      
\end{figure}

Moreover, we have identified other new transitions in $^{124}$Sn by analyzing 
the spectra 
in double coincidence with the 1047-keV $\gamma$ ray and one
transition of its main complementary fragments (either 
$^{116,118}$Cd or $^{96,98}$Zr). An example of coincidence spectrum
doubly-gated on some of the new transitions is shown in Fig.~\ref{yrast124Sn}. 
The resulting cascade resembles well those assigned to the lighter isotopes, but 
the link such as the 666-keV transition (in $^{122}$Sn) and the 782-kev one 
(in $^{120}$Sn), which is not observed any more. 

We have looked for a second branch directly linked to the 7$^-$ isomeric 
state, as measured in  $^{120,122}$Sn. In the spectrum gated by the 229-keV
line (SAPhIR experiment), a 252-keV transition is observed, as well as a weak
377-keV transition, these two transitions could be part of the foreseen cascade
ending to the 7$^-$ state. Unfortunately, no high energy transition could be
unambiguously seen in that spectrum, allowing us to complete the cascade. 
Nevertheless, using the Euroball data sets,
we identified a cascade of three transitions (253~keV, 377~keV, and 
1116~keV) which does belong to $^{124}$Sn since they are detected in
coincidence with the complementary fragments expected in the two fusion 
reactions, namely $^{116,118}$Cd and $^{96,98}$Zr. Then we
have assumed that the 253~kev transition measured in the present work is
the one decaying the 8$^+$ state towards the 7$^-$ isomeric
state~\cite{NNDC}. All these arguments would lead to a second decay path of the
4323~keV state, 252+377+1116+253. It is worth mentioning that the number of
counts of the 253-1116-377 events measured in the Euroball data sets is      
very high as compared to the very weak number of events containing the 
252-keV transition
located just above the 4071-keV state. This could be due to a large side feeding
of that state. Since some intensities of the coincidence events establishing
that second cascade are very weak, we have chosen to draw it with dotted lines
in Fig.~\ref{schema124}.

All the transitions newly observed  in $^{124}$Sn are given in 
Table~\ref{gammas124Sn}.
The spin and parity of the new states will be discussed and
assigned in Sec.~\ref{spin_assignment}.
\begin{table}[!h]
\caption{Properties of the new transitions assigned to $^{124}$Sn 
in this experiment. The energies of the two long-lived isomeric states
at 2325.0 keV (I$^\pi = 7^-$) and 2656.6 keV (I$^\pi = 10^+$), as well as this
of the 8$^+$ state, (written in bold) are from Ref.~\cite{NNDC}.}
\label{gammas124Sn}
\begin{ruledtabular}
\begin{tabular}{rrccc}
E$_\gamma$(keV)\footnotemark[1]&I$_\gamma$\footnotemark[1]$^,$\footnotemark[2]&J$_i^\pi$$\rightarrow$J$_f^\pi$  &E$_i$(keV)&E$_f$(keV)\\
\hline
228.5(4) &13(4) &  15$^-$ $\rightarrow$ 13$^-$   & 4551.8& 4323.2 \\
251.7(4) &4(2) & 13$^-$  $\rightarrow$ 11$^-$	& 4323.2& 4071.4  \\
253.2(3) &22(7) & 8$^+$  $\rightarrow$ 7$^-$	& 2578.4& {\bf 2325.0} \\
377.3(3) &7(3)  & 11$^-$  $\rightarrow$ 9$^-$	& 4071.4& 3694.1 \\
490.2(3) &39(6) & 16$^+$  $\rightarrow$ 14$^+$	& 5189.4& 4699.2\\
619.8(2) &43(7) & 13$^-$  $\rightarrow$ 12$^+$   & 4323.2& 3703.4\\
762.7(3) &25(5)  & (18$^+$) $\rightarrow$ 16$^+$  & 5952.1& 5189.4\\			
995.8(3) &57(9) &  14$^+$ $\rightarrow$ 12$^+$	& 4699.2& 3703.4\\
1024.3(4)&11(3) &  ~~~~ $\rightarrow$ (18$^+$)	& 6976.4& 5952.1\\
1046.8(3)&100 &  12$^+$ $\rightarrow$ 10$^+$   & 3703.4& {\bf 2656.6}\\
1115.7(3)&13(4)  & 9$^-$   $\rightarrow$ 8$^+$    & 3694.1& {\bf 2578.4}\\
1177.7(5)&5(2) & (17$^-$)$\rightarrow$ 15$^-$   & 5729.4& 4551.8\\
\end{tabular}
\end{ruledtabular}
\footnotetext[1]{The number in parenthesis is the error in the last digit.}
\footnotetext[2]{The relative intensities are normalized to the sum 
$I_\gamma(996) + I_\gamma(620)=100$.}
\end{table}

\subsubsection{$^{126}$Sn}
The two most intense transitions of the new cascade assigned to $^{126}$Sn have 
energies of 1030~keV and 570~keV (see Sec.~\ref{identification}).   
The doubly-gated spectrum exhibits new transitions at 180~keV, 1149~keV and
762~keV. 
\begin{figure}[!h]
\includegraphics*[height=12.8cm]{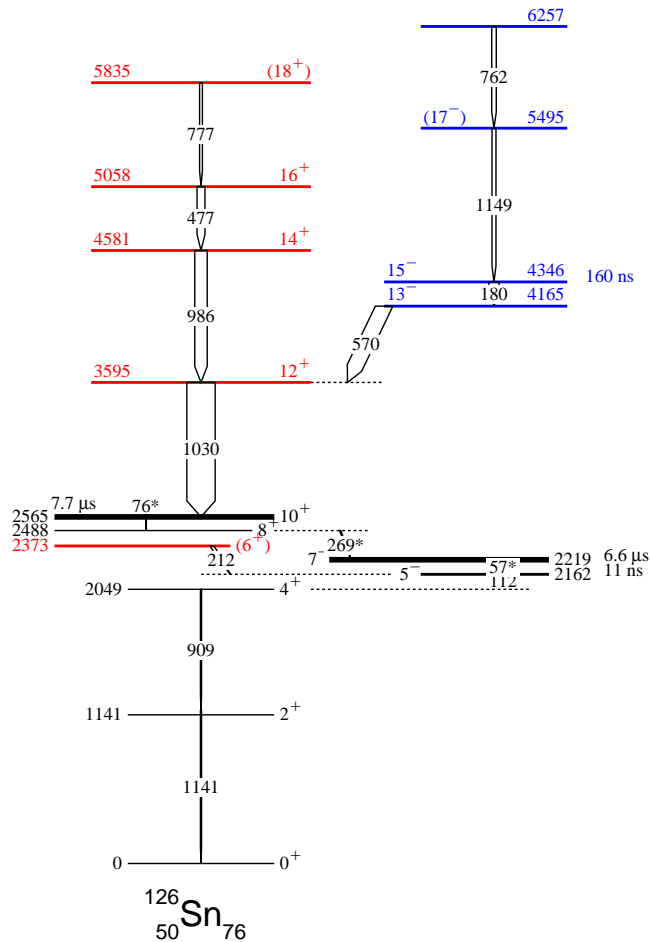}
\caption{(Color online) Level scheme of $^{126}$Sn deduced in the present work. The colored levels
are new.
The two long-lived isomeric 
states and their $\gamma$-decays to the low-lying states were already
known~\cite{NNDC}. The 269-keV transition, located between 
two long-lived isomeric states,  as well as the very converted
76- and 57-keV transitions, could not be observed in our work. 
The width of
the arrows is representative of the relative intensity of the $\gamma$ rays
above the isomeric states.  
}
\label{schema126}      
\end{figure}
Moreover the data of the SAPhIR experiment indicate that the
cascade comprising the first three transitions, at 1030~keV, 570~keV, and 
180~keV, is delayed. Using the same procedures as previously described, we
have measured the half-life, T$_{1/2}=$160(20)~ns. Moreover the spectrum
doubly-gated by the 1030-keV line and one transition of a complementary fragment
reveals other $\gamma$ rays belonging to $^{126}$Sn, such as 986~keV, 477~keV
and 777~keV, which form another cascade. Thus  
$^{126}$Sn exhibits two parallel structures having one transition in common,
1030~keV. In comparison to the results obtained in the lighter isotopes, we have
placed it directly above the 10$^+$ isomeric state (see Fig.~\ref{schema126}). 
Then we have looked for a second decay path of the new isomeric state ending 
to the 7$^-$ state at 2219~keV, but we have found no candidate in our data sets.
This is likely due to the lack of statistics, $^{126}$Sn being in the 
heavy-A tail of the Sn fragment distribution.  

\begin{figure}[!h]
\includegraphics*[width=8cm]{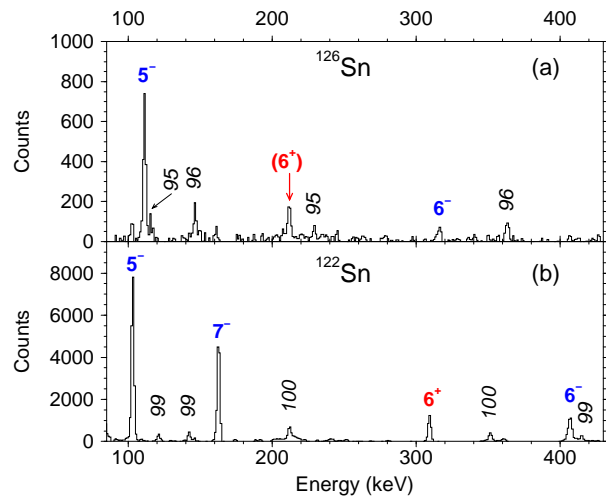}
\caption{(Color online) Coincidence spectra double-gated on the first two  
transitions of $^{126}$Sn (a) and $^{122}$Sn (b), built 
from the $^{18}$O + $^{208}$Pb data set. The $\gamma$-rays emitted
by $^{126,122}$Sn are labelled by the spin and
parity values of the decaying states, and those emitted by the Zr 
complementary fragments are labelled 
by their masses A, written in italics.  
}
\label{lesixplus}      
\end{figure}

Up to now, no 6$^+$ state was measured in $^{126}$Sn,
whilst such a state is expected to lie below the 8$^+$ state, as in
the other even-$A$ isotopes. The E2 decays of the 8$^+_1$ and 6$^+_1$ states are
hindered because of their low energy, thus these levels decay by means of E1
transitions towards negative parity states, 7$^-$ and 5$^-$ respectively (see
Figs.~\ref{schema122} and \ref{schema124}). In the present work, we have observed the  
6$^+_1$ $\rightarrow$ 5$^-$ transitions of $^{120,122,124}$Sn
in coincidence with the first three transitions of their level schemes. 
Thus we have looked for the  6$^+_1$ $\rightarrow$ 5$^-$ $\gamma$ ray 
in $^{126}$Sn, by analyzing the spectra doubly-gated by its first transitions
(at 1141 keV, 909 keV, and 112 keV). Figure~\ref{lesixplus} shows two spectra, 
built from the $^{18}$O + $^{208}$Pb data set and
using similar conditions for $^{126}$Sn and $^{122}$Sn. All the observed 
$\gamma$ lines are assigned to the decay of known states of either 
$^{126}$Sn ($^{122}$Sn) or their Zr complementary fragments, except the  
new $\gamma$ line at 212~keV, which is assigned to the
decay of the 6$^+_1$ state of $^{126}$Sn (see Fig.~\ref{schema126}). 

Noteworthy is the fact that 212~keV is also the energy of the 
2$^+$ $\rightarrow$ 0$^+$ transition of $^{100}$Zr. The observation of prompt
coincidence between a $\gamma$ ray at this energy and the first transitions of 
$^{126}$Sn could have been interpreted as the fact that $^{100}$Zr is a
complementary fragment of $^{126}$Sn, meaning that the compound nucleus of the 
$^{18}$O + $^{208}$Pb reaction, $^{226}$Th, may fission before emitting 
any neutron. Such process has never been observed, as expected since neutron 
emission from an excited compound nucleus is always faster than the 
fission. The coincidence relationships measured in the 
$^{12}$C + $^{238}$U data set corroborate the 
location of the 212-keV $\gamma$ line in the $^{126}$Sn level scheme, which
rules out a misinterpretation of the 212-keV $\gamma$ line.

All the transitions newly observed  in $^{126}$Sn are given in 
Table~\ref{gammas126Sn}.
The spin and parity of the new states will be discussed and
assigned in Sec.~\ref{spin_assignment}.  
\begin{table}[!h]
\caption{Properties of the new transitions assigned to $^{126}$Sn 
in this experiment. The energies of the long-lived isomeric state
at 2564.5 keV (I$^\pi = 10^+$)and
of the 5$^-$ state at 2161.5 keV (written in bold) are from Ref.~\cite{NNDC}.}
\label{gammas126Sn}
\begin{ruledtabular}
\begin{tabular}{rrccc}
E$_\gamma$(keV)\footnotemark[1]&I$_\gamma$\footnotemark[1]$^,$\footnotemark[2]&J$_i^\pi$$\rightarrow$J$_f^\pi$  &E$_i$(keV)&E$_f$(keV)\\
\hline
180.5(3) & 36(7) &  15$^-$ $\rightarrow$ 13$^-$   & 4345.7& 4165.2 \\
211.7(4) & 12(4) & (6$^+$)  $\rightarrow$ 5$^-$	  & 2373.2& {\bf 2161.5} \\
476.7(3) & 28(7) & 16$^+$  $\rightarrow$ 14$^+$	  & 5057.9& 4581.2\\
570.5(3) & 56(11)& 13$^-$  $\rightarrow$ 12$^+$   & 4165.2& 3594.7\\
761.7(4) & 8(3)  &  ~~~~ $\rightarrow$ (17$^-$)	  & 6256.9& 5495.2\\
777.1(4) & 9(3)  & (18$^+$) $\rightarrow$ 16$^+$  & 5835.0& 5057.9\\			
986.5(3) & 44(9) &  14$^+$ $\rightarrow$ 12$^+$	  & 4581.2& 3594.7\\
1030.2(3)&100    &  12$^+$ $\rightarrow$ 10$^+$   & 3594.7& {\bf 2564.5}\\
1149.5(5)& 13(4) & (17$^-$)$\rightarrow$ 15$^-$   & 5495.2& 4345.7\\
\end{tabular}
\end{ruledtabular}
\footnotetext[1]{The number in parenthesis is the error in the last digit.}
\footnotetext[2]{The relative intensities are normalized to the sum 
$I_\gamma(986) + I_\gamma(570)=100$.}
\end{table}
\subsubsection{Angular momentum and parity values of the high-spin states of 
$^{120-126}$Sn}\label{spin_assignment}

Given that the new high-spin structures of $^{120-126}$Sn reported in the 
previous sections are very close to each other, we assume that the similar 
states in the four level schemes have the same spin and parity values.

First, we have extracted the internal conversion coefficients of the isomeric
transitions of $^{120-126}$Sn (at 241 keV, 242 keV, 229 keV, and 180 keV
respectively) by analyzing the relative intensities of transitions in cascade. 
The intensity imbalances of the 241-, 242-, and 229-keV $\gamma$ rays measured
in spectra in double coincidence with at least one transition located above them
in their respective level scheme lead to $\alpha_{tot} \le 0.1$.  
This is consistent with $E1$, $M1$, or $E2$ assignment. On the other hand, 
the intensity imbalance of the 180-keV $\gamma$ ray of $^{126}$Sn gives 
$\alpha_{tot} = 0.25(5)$, in good agreement with the theoretical value for $E2$
multipolarity, $\alpha_{tot}(E2, 180$~keV) = 0.20~\cite{BRICC}. 
Assuming that the nature of the former transitions are also $E2$, we have calculated
all the $B(E2)$ values, which are reported in Table~\ref{BE2}. One has to note
that these $B(E2)$ values have the same order of magnitude as those already
measured for the isomeric decay of some lower energy states, either in the even-$A$ isotopes or
in the odd-A ones. This will be discussed in Sec.~\ref{discuss}. 
\begin{table}[!h]
\caption{Properties of the new isomeric states of $^{120-126}$Sn}
\label{BE2}
\begin{ruledtabular}
\begin{tabular}{cccccc}
Nucleus & $E_i$&$E_\gamma$& T$_{1/2}$\footnotemark[1]&
$B(E2)$\footnotemark[1]&$B(E2)$\footnotemark[1]\\
	& keV & keV & ns & $e^2fm^4$ & W.u.\\
\hline
$^{120}$Sn &4890.1 &241.1 &36(4)  & 18(2) & 0.51(5)\\
$^{122}$Sn &4720.5 &242.1 &146(15) & 4.4(4) & 0.12(1)\\
$^{124}$Sn &4551.8 &228.5 &260(25) & 3.2(3) & 0.09(1)\\
$^{126}$Sn &4345.7 &180.5 &160(20) & 16(2) & 0.42(5)\\
\end{tabular}
\end{ruledtabular}
\footnotetext[1]{The number in parenthesis is the error in the last digit.}
\end{table}

Secondly, we have analyzed the $\gamma-\gamma$ angular correlations of 
strongest transitions. The experimental results are given in 
Table~\ref{correl_Sn}. 
\begin{table}[!h]
\caption{Coincidence rates between $\gamma$-rays of $^{120-124}$Sn 
as a function of their relative angle of detection, normalized to 
the ones obtained around 75$^\circ$.
}
\label{correl_Sn}
\begin{ruledtabular}
\begin{tabular}{cccccc}
&&E$_\gamma$-E$_\gamma$&R(22$^\circ$)\footnotemark[1]&R(46$^\circ$)\footnotemark[1]
&R(75$^\circ$)\footnotemark[1]\\
\hline

&$^{120}$Sn  &1190~-~557  &  0.87(9)& 0.95(7)  &1.00(5) \\
&            &1190~-~241  &  1.1(1)& 1.12(8)   &1.00(5) \\
& &		&	&		&		\\		
&$^{122}$Sn  &1103~-~1030  &  1.17(9) & 1.07(7)&1.00(5) \\
&            &1103~-~609  &  0.8(1) & 0.96(7)  &1.00(5)	\\
&            &1103~-~242  &  1.2(1) & 1.14(8)  &1.00(5)	\\
& &		&	&		&		\\
&$^{124}$Sn  &1047~-~996  &  1.15(9) & 1.09(7) &1.00(5)	\\
&            &1047~-~490  &  1.06(9) & 1.11(7) &1.00(5)	\\
&            &996~-~490  &  1.2(1) & 1.05(7)   &1.00(5) \\
\end{tabular}
\end{ruledtabular}
\footnotetext[1]{The number in parenthesis is the error in the last digit.}
\end{table}
The coincidence rates measured for 6 pairs indicate that all their 
transitions have the same multipole order: The 1190- and 241-keV transitions (in
$^{120}$Sn), the  1103-, 1030-, and 242-keV transitions (in $^{122}$Sn), 
the 1047-, 996-, and 490-keV transitions (in $^{124}$Sn). Since the 241- and
242-keV transitions are $E2$, a quadrupole order is also assigned to the 
1190-, 1103-, and 1030-keV transitions. Then the 1047-keV $\gamma$ ray, located
just above the 10$^+$ state of $^{124}$Sn is also $E2$, 
as the 1190- and  1103-keV $\gamma$
rays in $^{120}$Sn and $^{122}$Sn, respectively . 
On the other hand, 
both the 557- and the 609-keV transitions have a different multipole order 
from that of the 1190- and  1103-keV transitions, respectively (see 
Table~\ref{correl_Sn}). Thus they are dipole transitions.

In conclusion, the cascades of three $E2$ transitions, placed
above the 10$^+$ isomeric states of $^{120-126}$Sn, define the 12$^+$, 14$^+$,
and 16$^+$ levels (see Figs.~\ref{schema120}, \ref{schema122}, \ref{schema124},
and~\ref{schema126}).
Moreover the second cascade identifed in each level scheme comprises 
states with odd spin values, the $I=13$ state decays to the 12$^+$ state and 
the $I=15$
state is isomeric. A negative parity is assigned to the states of this second
cascade since in $^{120,122}$Sn, the $I=13$ state is linked to the long-lived 7$^-$ state by means of
a cascade of three transitions. This leads to the 9$^-$ and 
11$^-$ states. All these assignments are reported in the level schemes drawn in
Figs.~\ref{schema120}, \ref{schema122}, \ref{schema124}, and~\ref{schema126}.
\subsubsection{Comparison with other results recently published}
During the completion of this work,
the decay of a new isomeric state in $^{128}$Sn populated in the fragmentation of 
$^{136}$Xe was reported~\cite{pi11}. Its deexcitation is very similar to that
of the isomeric states we have measured in $^{120-126}$Sn. The spin and parity
values of all the states involved in the isomeric decay were proposed by
comparison with results of shell-model calculations. 

Moreover, at the very end of the writing of this paper, we became acquainted with 
a publication on the high-spin states of the even-$A$ $^{118-124}$Sn~\cite{fo11}. 
The authors have used the fusion-fission process to populate a few levels lying 
above the long-lived 10$^+$ states, the identification of the first transition of 
each $\gamma$-ray cascade being confirmed thanks to the behavior of its 
excitation function in 
the $^{124}$Sn($n,xn\gamma$) reactions. The resulting four(three) new states of 
$^{120,124}$Sn($^{122}$Sn) are part of the structures identified in the present 
work.

\subsection{Study of the odd-A $^{}$Sn isotopes}\label{masseimpaire}
Populated in deep inelastic reactions, a few medium-spin levels were 
identified in the four odd-A $^{119-125}$Sn isotopes
prior to this work~\cite{ma94,zh00}. The maximum spin values measured in these isotopes were 27/2
in case of negative parity and 19/2 or 23/2 in case of positive parity. Such
states correspond to the breaking of one neutron pair in the $\nu h_{11/2}$
subshell which gives rise to several sets of states, depending on the subshell
occupied by the odd neutron. The $I^\pi = 27/2^-$ state, attributed to the 
$(\nu h_{11/2})^3$ configuration, is located between 3.1~MeV and 2.6~MeV. It is 
a long-lived isomeric state in $^{123}$Sn (34 $\mu s$), while the half-lives of
the 27/2$^-$ states of $^{119,121,125}$Sn isotopes are in a 30-230 ns range, i.e.
short enough such that coincidence events may be detected across the isomeric
states. The
maximum spin of the $(\nu h_{11/2})^2(\nu d_{3/2})$ configuration is 23/2$^+$.
Such a state, which is isomeric, is only known in $^{123,125}$Sn. On the other 
hand, the four isotopes exhibit a long-lived isomeric 19/2$^+$ state, from 
the $(\nu h_{11/2})^2(\nu s_{1/2})$ configuration, located around 2~MeV 
excitation energy. 

Using the data of the two fusion-fission reactions of the present work, we have
identified, for the first time, several structures of $^{119-125}$Sn located
above 3~MeV excitation energy and spin values greater than 27/2. In the 
following, we first present the building of each high-spin level scheme. 
Then, we discuss the angular momentum and 
parity assignments of most of the new states of $^{119-125}$Sn. 
  
\subsubsection{$^{119}$Sn and $^{125}$Sn}\label{119et125}
The high-spin states of $^{119}$Sn and $^{125}$Sn are weakly populated in 
the two fusion-fission reactions used in the present work, as these two isotopes 
are in the tails of the Sn fragment distribution. The half-live of their 
27/2$^-$ states (at 3101~keV and 2624~keV respectively) are short
enough to let the detection of coincidence events across the isomers. 
Therefore the search for $\gamma$-rays populating their 27/2$^-$ states is
more appropriate in the spectra gated by their low-lying transitions than in the spectra
gated by their complementary fragments (namely $^{120-122}$Cd for $^{119}$Sn and 
$^{114-116}$Cd for $^{125}$Sn).

\begin{figure}[!h]
\includegraphics*[height=9.7cm]{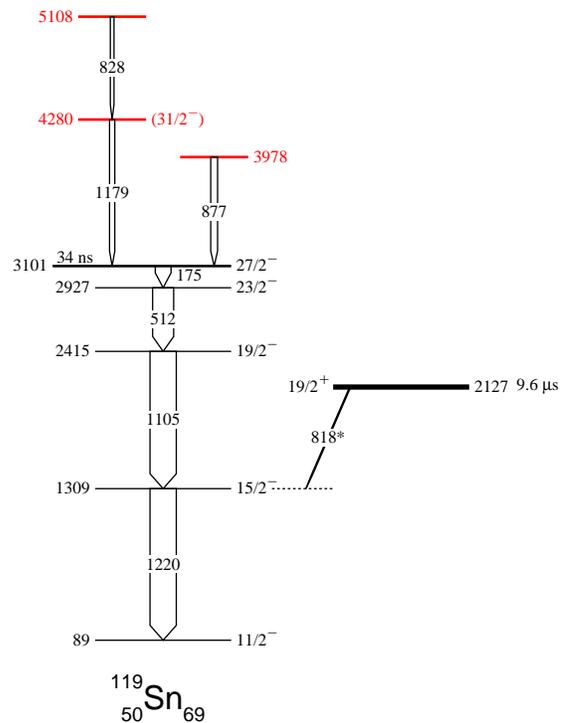}
\caption{(Color online) Level scheme of $^{119}$Sn deduced in the present work. The colored levels
are new. The width of
the arrows is representative of the relative intensity of the $\gamma$ rays.
The two isomeric states were already known~\cite{NNDC}. In our work, the 818-keV transition could not 
be observed, nevertheless the 19/2$^+$ isomeric state is drawn for the sake of 
completeness. 
}
\label{schema119}      
\end{figure}

\begin{figure}[!h]
\includegraphics*[height=10cm]{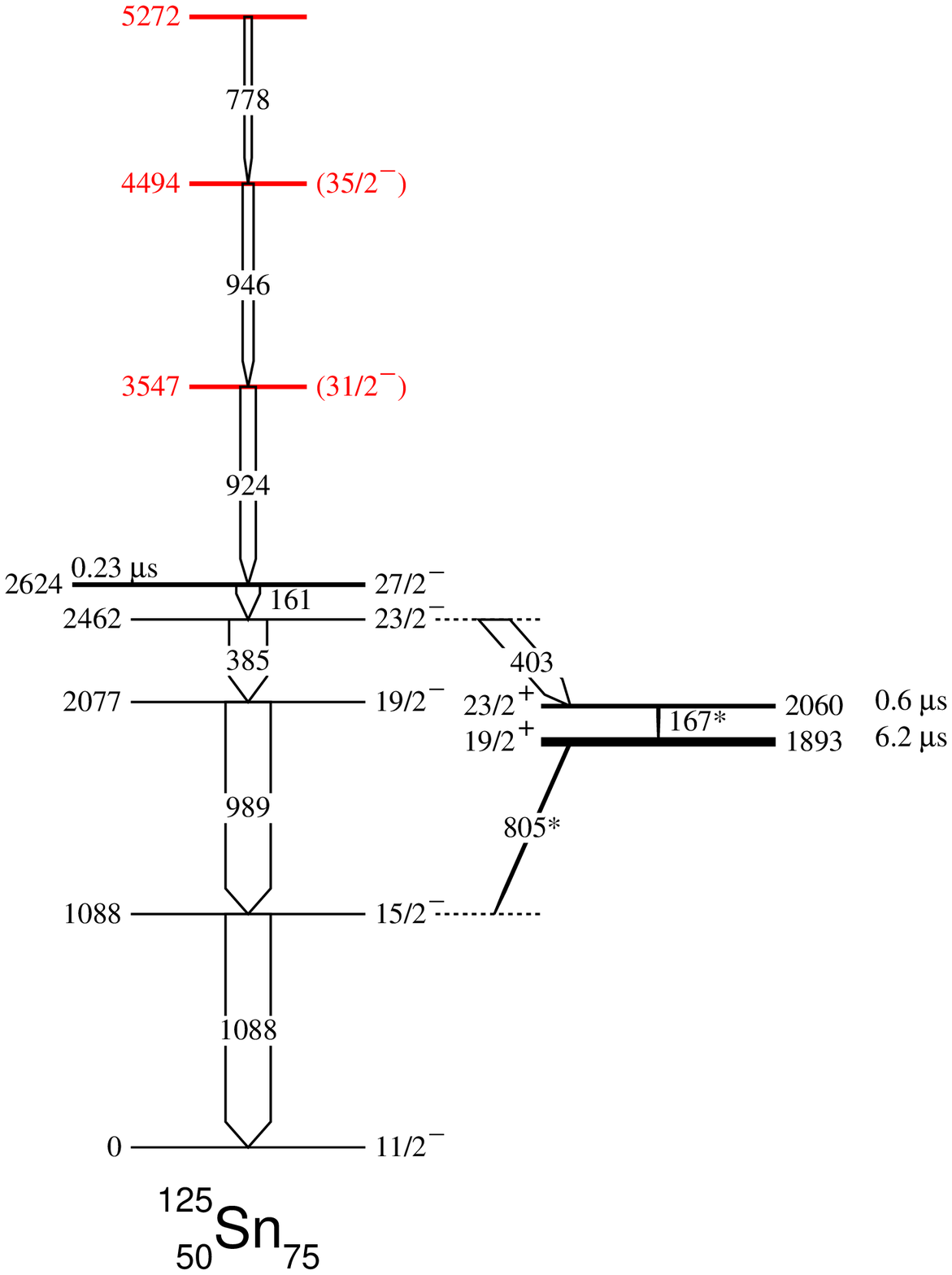}
\caption{(Color online) Level scheme of $^{125}$Sn deduced in the present work. The colored levels
are new. The width of
the arrows is representative of the relative intensity of the $\gamma$ rays.
The three isomeric states were already known~\cite{NNDC}. 
In our work, the 805-keV transition could not be observed and the very converted
167-keV line was extremely weak because of the lifetime of the decaying 
state.
}
\label{schema125}      
\end{figure}

In both data sets registered in our work, all the
spectra doubly-gated by the $\gamma$- transitions decaying the 27/2$^-$ state of 
$^{119}$Sn reveal new $\gamma$-lines which have to be placed above it 
(see Fig.~\ref{schema119}). 
As for $^{125}$Sn, a cascade of three new
$\gamma$- transitions is observed in coincidence with the transitions decaying the 
27/2$^-$ state at 2624 keV (see Fig.~\ref{schema125}). The new transitions of 
$^{119,125}$Sn are also observed in spectra gated by $\gamma$ rays emitted by
their Cd partners.
All the transitions observed  in $^{119,125}$Sn are given in 
Tables~\ref{gammas119Sn} and \ref{gammas125Sn}, respectively.
The spin and parity of the new states will be discussed 
in Sec.~\ref{assign_spin_oddA}.  
\begin{table}[!h]
\caption{Properties of the $^{119}$Sn transitions. The excitation 
energy of the 11/2$^-$ state  (written in bold) is
from Ref.~\cite{NNDC}}
\label{gammas119Sn}
\begin{ruledtabular}
\begin{tabular}{rrccc}
E$_\gamma$(keV)\footnotemark[1]&I$_\gamma$\footnotemark[1]$^,$\footnotemark[2]&J$_i^\pi$$\rightarrow$J$_f^\pi$  &E$_i$(keV)&E$_f$(keV)\\
\hline
174.6(3) & 61(9) &  27/2$^-$ $\rightarrow$ 23/2$^-$   & 3101.1 & 2926.6\\
511.9(3) & 80(12) & 23/2$^-$ $\rightarrow$ 19/2$^-$    & 2926.6 & 2414.7 \\
827.8(4) & 15(5) &  				    & 5107.9 & 4280.1\\
876.8(4) & 26(6) & ~~~~~~  $\rightarrow$ 27/2$^-$     & 3977.9 & 3101.1\\
1105.2(3) & 100 &  19/2$^-$ $\rightarrow$ 15/2$^-$   & 2414.7 & 1309.5 \\
1179.0(5) & 21(5) &  ~~~~~~ $\rightarrow$ 27/2$^-$    & 4280.1 & 3101.1\\			
1220.0(3) & 100 &  15/2$^-$ $\rightarrow$ 11/2$^-$   & 1309.5 & {\bf 89.5}\\
\end{tabular}
\end{ruledtabular}
\footnotetext[1]{The number in parenthesis is the error in the last digit.}
\footnotetext[2]{The relative intensities are normalized to 
$I_\gamma(1105) =100$.}
\end{table}
\begin{table}[!h]
\caption{Properties of the $^{125}$Sn transitions.}
\label{gammas125Sn}
\begin{ruledtabular}
\begin{tabular}{rrccc}
E$_\gamma$(keV)\footnotemark[1]&I$_\gamma$\footnotemark[1]$^,$\footnotemark[2]&J$_i^\pi$$\rightarrow$J$_f^\pi$  &E$_i$(keV)&E$_f$(keV)\\
\hline
161.3(3) & 51(8) &  27/2$^-$ $\rightarrow$ 23/2$^-$   &2623.3 & 2462.0\\
385.6(3) & 84(12) & 23/2$^-$ $\rightarrow$ 19/2$^-$   & 2462.0&2076.4 \\
402.8(4) & 52(12) & 23/2$^-$ $\rightarrow$ 23/2$^+$   & 2462.0&2059.2 \\
778.4(5) & 16(5) &  				& 5271.9& 4493.5\\
923.7(4) & 34(8) & ~~~~~~  $\rightarrow$ 27/2$^-$   &3547.0 & 2623.3\\
946.5(4) & 24(7) &  				  &4493.5 & 3547.5\\
988.7(3) & 100 &  19/2$^-$ $\rightarrow$ 15/2$^-$   & 2076.4&1087.7 \\
1087.7(3) & 100 &  15/2$^-$ $\rightarrow$ 11/2$^-$   & 1087.7& 0\\
\end{tabular}
\end{ruledtabular}
\footnotetext[1]{The number in parenthesis is the error in the last digit.}
\footnotetext[2]{The relative intensities are normalized to 
$I_\gamma(989) =100$.}
\end{table}

We have not identified any $\gamma$-ray cascade which would 
be placed above the 19/2$^+$ or 23/2$^+$ states of $^{119}$Sn or $^{125}$Sn.
Because of the long lifetime of these isomeric states, the transitions located 
above them can be only identified from their coincidences with transitions 
emitted by the partners. 
When taking into account that the weak population of these two Sn isotopes is 
shared between several complementary fragments, every coincidence rate 
was too low in the present work.  

\subsubsection{$^{121}$Sn}\label{$^{121}$Sn}

\begin{figure}[!h]
\includegraphics*[width=7.5cm]{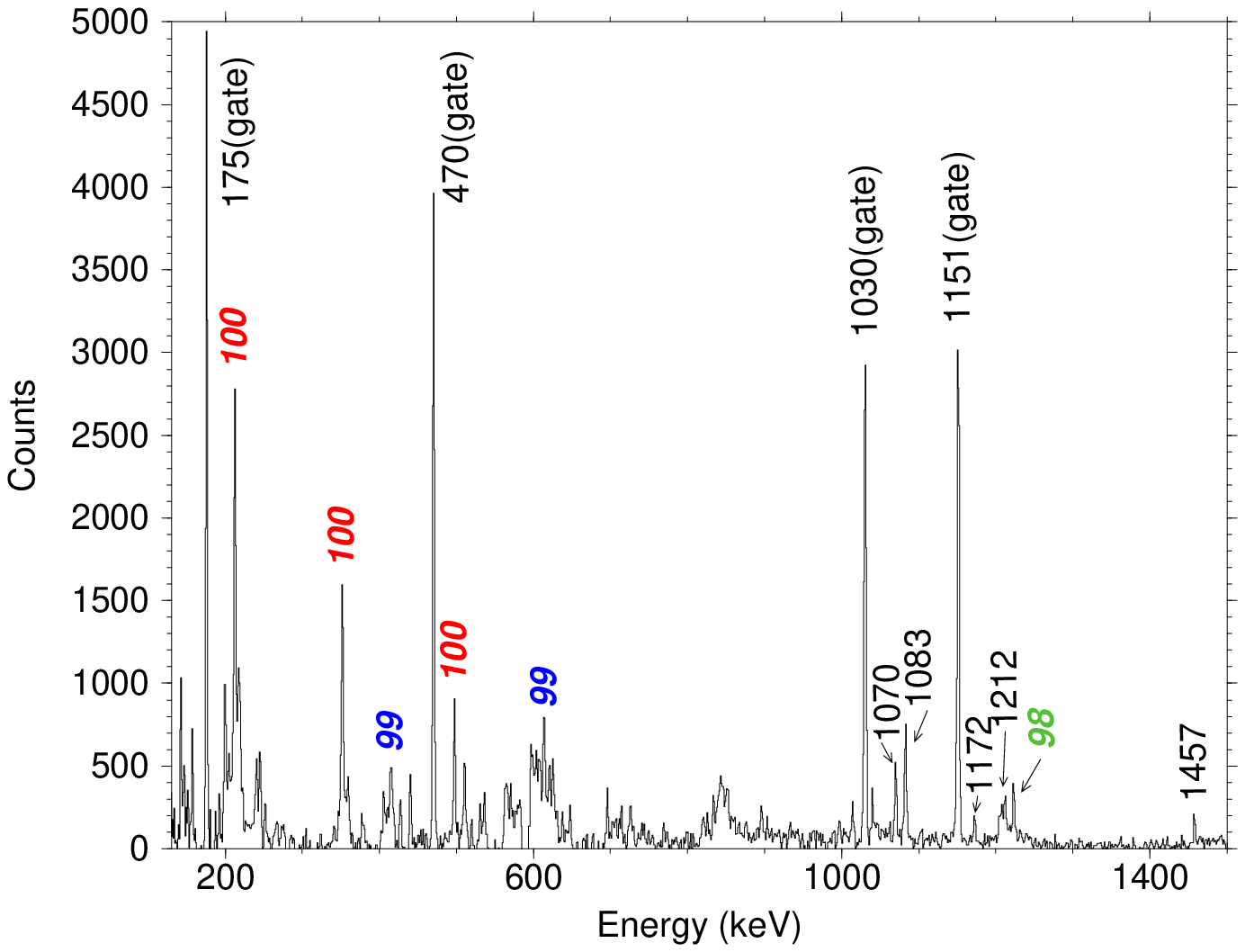}
\caption{(Color online) Coincidence spectrum double-gated on the 1151-, 1030-, 470- and
175-keV transitions of the yrast cascade built on the low-lying
11/2$^-$ state of $^{121}$Sn produced in the $^{18}$O + $^{208}$Pb 
reaction. 
The $\gamma$-rays emitted by the Zr complementary fragments are 
labelled by their masses A, written in italics.  
}
\label{spectre121Sn}      
\end{figure}

\begin{figure}[!h]
\includegraphics*[height=13.4cm]{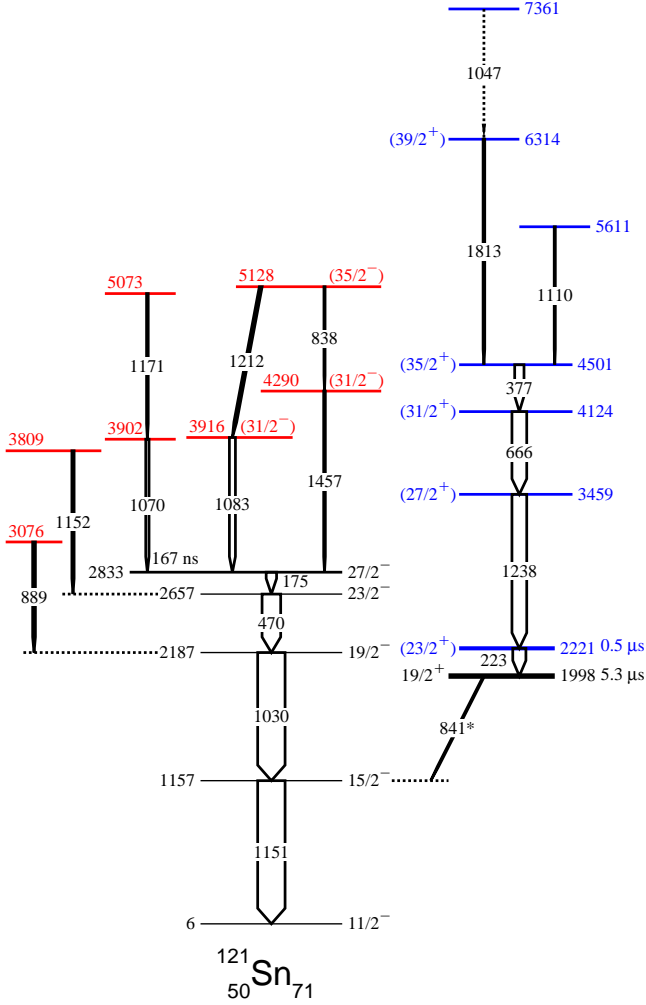}
\caption{(Color online) Level scheme of $^{121}$Sn deduced in the present work. The colored levels
are new. The isomeric 
27/2$^-$ and 19/2$^+$ states and their decays were already known~\cite{NNDC}. 
The 841 keV transition could not be observed in our work. The width of
the arrows is representative of the relative intensity of the $\gamma$ rays.
}
\label{schema121}      
\end{figure}

Prior to this work, the knowledge of the medium spin states of $^{121}$Sn was
very similar to the one of $^{119}$Sn, namely the 27/2$^-$ and 19/2$^+$ states 
and their decay towards the yrast levels having lower spin values. As in 
$^{119,125}$Sn, the half-life of the 27/2$^-$ isomer is short
enough to measure the coincidence events across it. An example of coincidence 
spectrum double-gated on the 1151-, 1030-, 470-, and
175-keV transitions of the yrast cascade is drawn in Fig.~\ref{spectre121Sn}.
Besides the transitions emitted by the complementary fragments, $^{100-98}$Zr,
it exhibits new peaks at 1070, 1083, 1171, 1212, and 1457~keV. In a second step,
we have analyzed the coincidence relationships of these new $\gamma$-lines 
and placed all these transitions in three cascades above the 
27/2$^-$ state at 2833~keV (see Fig.~\ref{schema121}). In addition two other
lines were observed in coincidence with a part of the yrast transitions, thus
the 889 keV transition defines a new state at 3076 keV and the 1152 keV one a
new state at 3809 keV. 

\begin{figure}[!h]
\includegraphics*[width=8cm]{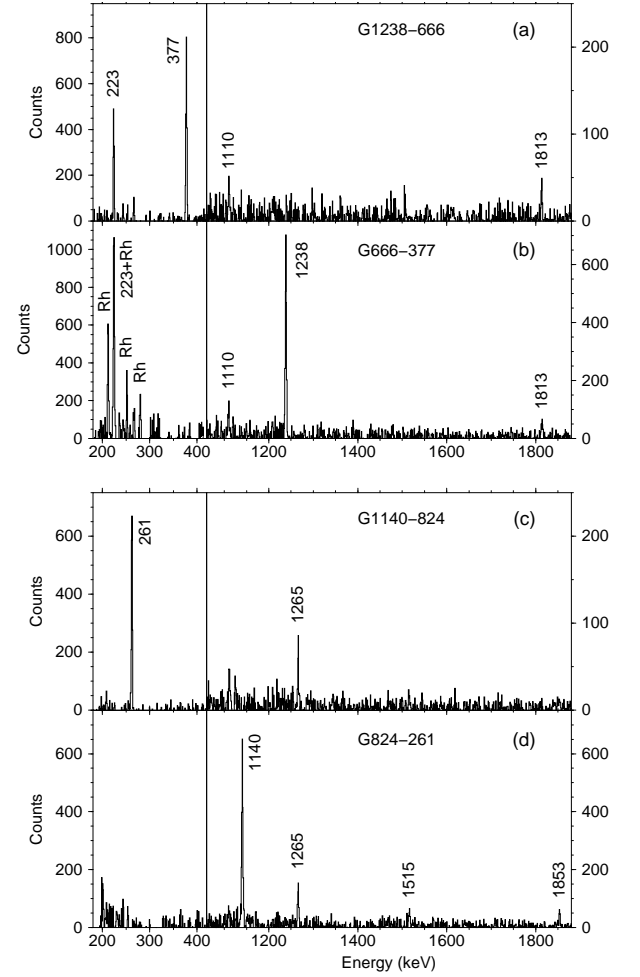}
\caption{Coincidence spectra double-gated on transitions belonging
to two new cascades, built from the $^{12}$C + $^{238}$U data set, in two energy
ranges, [180-420]~keV and [1060-1880]~keV. 
(a) and (b) Cascade emitted by $^{121}$Sn. The lines labelled by Rh are pollutions
(they belong to $^{111}$Rh, the 667- and 378-keV transitions being part of its
yrast cascade~\cite{ve02}).
(c) and (d) Cascade emitted by $^{123}$Sn. 
}
\label{sp121Snet123Sn}      
\end{figure}

The two most intense transitions of the new cascade assigned to $^{121}$Sn thanks
to their coincidences with their complementary fragments (see
Sec.~\ref{identification}) have energies of 1238 and 666~keV. In low-energy
part of the double gate set on these two transitions (see  
Fig.~\ref{sp121Snet123Sn}(a)), two new transitions, at 223 keV and 377 kev, are
clearly observed, the intensity of the 223 keV peak being the lowest.  
This would indicate that the 223 keV transition has to be located at the  
top of the cascade. 
Nevertheless, because of the comparison with the new cascade observed in
$^{123}$Sn (see Sec.~\ref{$^{123}$Sn}), we have chosen to put this transition 
at the bottom of the cascade, assuming that it deexcites an isomeric state with 
a half-life long enough to lower its relative intensity. The measured imbalance 
leads to T$_{1/2} = 0.5(1) \mu s$ when taking into account 
the conversion coefficient for an $E2$ transition, $\alpha_{tot}(E2,$223 keV)=
0.096~\cite{BRICC}. Unfortunately, the data from the SAPhIR experiment cannot be
used to determine the half-life of this isomeric state, as it only emits one
transition (in the time window of Euroball). The corresponding events contain a
unique $\gamma$-ray, this is not enough to select unambiguously the emitting
nucleus. 

In the high-energy part of the spectrum shown in Fig.~\ref{sp121Snet123Sn}(a), we 
observe two
weak transitions, at 1110 and 1813 keV, which are also present in the spectrum
doubly-gated by the 666- and 377-keV transitions 
(see Fig.~\ref{sp121Snet123Sn}(b)). They are located at the top of 
the new cascade. 
Finally, the whole set is placed above 
the long-lived isomeric state at 1998 keV, which is the only solution to explain
why these transitions do belong to the level scheme of  $^{121}$Sn while
they are not detected in coincidence with its well-known yrast transitions.

All the transitions observed  in $^{121}$Sn are given in 
Table~\ref{gammas121Sn}. The spin and parity of the new states will be 
discussed in Sec.~\ref{assign_spin_oddA}.  
\begin{table}[!h]
\caption{Properties of the $^{121}$Sn transitions. The excitation 
energies of the 11/2$^-$ and 19/2$^+$ states  (written in bold) are
from Ref.~\cite{NNDC}}
\label{gammas121Sn}
\begin{ruledtabular}
\begin{tabular}{rrccc}
E$_\gamma$(keV)\footnotemark[1]&I$_\gamma$\footnotemark[1]$^,$\footnotemark[2]&J$_i^\pi$$\rightarrow$J$_f^\pi$  &E$_i$(keV)&E$_f$(keV)\\
\hline
175.4(2)  & 38(9) &  27/2$^-$ $\rightarrow$ 23/2$^-$    & 2832.7 & 2657.3\\
222.8(3)  & 48(10)\footnotemark[3] & (23/2$^+$) $\rightarrow$ 19/2$^+$   & 2220.8 & {\bf 1998.0}\\
376.9(3)  & 33(7) & (35/2$^+$) $\rightarrow$ (31/2$^+$) & 4501.2 &  4124.3\\
470.2(2)  & 68(13) &  23/2$^-$ $\rightarrow$ 19/2$^-$    & 2657.3 & 2187.1 \\
665.7(3)  & 53(11) & (31/2$^+$) $\rightarrow$ (27/2$^+$) & 4124.3 &  3458.6\\
837.8(4)  & 3(1) &(35/2$^-$) $\rightarrow$ (31/2$^-$)  & 5128.0 & 4290.1 \\ 
889.4(4)  & 9(3) &  ~~~~~~  $\rightarrow$ 19/2$^-$     & 3076.5 & 2187.1\\
1029.8(3) & 100 &  19/2$^-$ $\rightarrow$ 15/2$^-$   & 2187.1 &  1157.3\\
1047.1(5) & weak &~~~~~~  $\rightarrow$ (39/2$^+$)   & 7361.3 & 6314.2\\
1069.7(4) & 12(4) & ~~~~~~  $\rightarrow$ 27/2$^-$      & 3902.4 & 2832.7\\
1082.9(4) & 21(4) & (31/2$^-$)$\rightarrow$ 27/2$^-$     & 3915.6 & 2832.7\\			
1109.9(5) & 4(3) &  ~~~~~~  $\rightarrow$ (35/2$^+$) & 5611.1 & 4501.2\\
1151.0(3) & 100 &  15/2$^-$ $\rightarrow$ 11/2$^-$   & 1157.3 & {\bf 6.3}\\
1151.6(4) & 8(4) &  ~~~~~~  $\rightarrow$ 23/2$^-$     & 3808.9 & 2657.3\\
1170.6(5) & 6(3) &  				    & 5073.0 & 3902.4\\
1212.5(4) & 9(3) & (35/2$^-$) $\rightarrow$ (31/2$^-$)    & 5128.0 & 3915.6\\
1237.8(3) & 53(11) & (27/2$^+$) $\rightarrow$ (23/2$^+$)  & 3458.6 & 2220.8\\
1457.4(4) & 6(3) &  (31/2$^-$) $\rightarrow$ 27/2$^-$ & 4290.1 & 2832.7 \\
1813.0(4) & 5(2) & (39/2$^+$) $\rightarrow$ (35/2$^+$)  & 6314.2 & 4501.2 \\
\end{tabular}
\end{ruledtabular}
\footnotetext[1]{The number in parenthesis is the error in the last digit.}
\footnotetext[2]{The relative intensities are normalized to 
$I_\gamma(1030) =100$.}
\footnotetext[3]{See text.}
\end{table}

\subsubsection{$^{123}$Sn}\label{$^{123}$Sn}

The 23/2$^+$ and 27/2$^-$ isomeric states were identified in $^{123}$Sn, their 
very long half-lives being
mainly due to the half-filling of the $\nu h_{11/2}$ subshell. Thus the
identification of all its excited states with I$^\pi > 27/2^-$ or 23/2$^+$ relies
on the coincidences with $\gamma$-rays emitted by its complementary fragments. 

As for the previous cases, we start with the two coincident transitions 
at 1140 and 824~keV, assigned to $^{123}$Sn because of the mass distribution 
of the complementary fragments in the two fusion-fission reactions 
(see Sec.~\ref{identification}). The low-energy part of the spectrum doubly-gated 
on these two transitions reveals one new $\gamma$-line at 261~keV 
(see Fig.~\ref{sp121Snet123Sn}(c)) and the high-energy part of the spectrum 
doubly-gated on the 261- and 824-keV transitions shows new high-energy low-intensity 
transitions (see Fig.~\ref{sp121Snet123Sn}(d)). 

\begin{figure}[!h]
\includegraphics*[height=14.6cm]{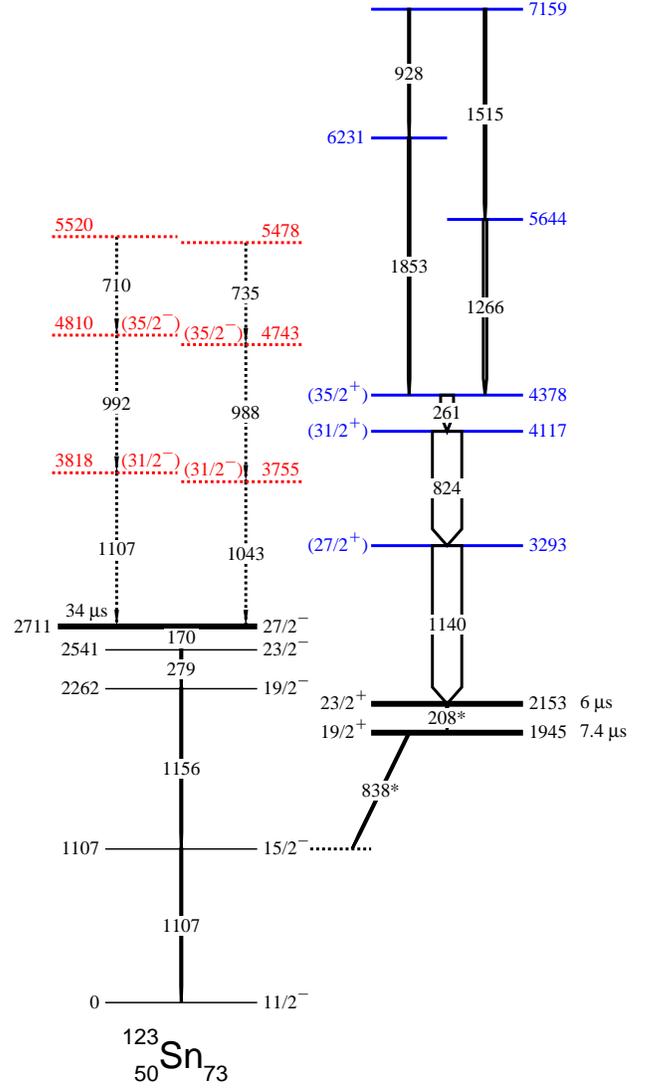}
\caption{(Color online) Level scheme of $^{123}$Sn deduced in the present work. The colored levels
are new. The isomeric 
27/2$^-$, 23/2$^+$ and 19/2$^+$ states and their decays were already 
known~\cite{NNDC}. 
The 208 and 838 keV transitions could not be observed in our work. as well as several other
decays of the 27/2$^-$ and 23/2$^-$ states~\cite{ma94}. Two new cascades are
tentatively placed above the 27/2$^-$ state (see text). The width of
the arrows is representative of the relative intensity of the $\gamma$ rays, except for that
of the decay of the 27/2$^-$ long-lived isomeric state (see text).
}
\label{schema123}      
\end{figure}

The resulting structure, which is very similar to the one assigned to $^{121}$Sn, 
is placed above the positive-parity isomeric state, the 23/2$^+$ at
2153~keV (see Fig.~\ref{schema123}). It is worth noting 
that, due to the time range defining the coincidence events (about 300 ns), 
the already-known E2 transition (208~keV) deexciting this long-lived state, T$_{1/2}=6~\mu$s,  
is never included in an event registered in our experiment. This explains why the new
structure of $^{123}$Sn contains one low-energy 
transition (261~keV) instead of the two transitions (223 and 377~keV) observed in
$^{121}$Sn, as discussed in the previous section.  
Noteworthy is the fact that the three transitions decaying the 4378-keV state of
$^{123}$Sn have not been measured in the SAPhIR experiment. Thus the half-life of 
this state is smaller than 30~ns.  

In addition, several weak transitions were observed in coincidence with those
emitted by the same partners as $^{123}$Sn (namely $^{116,118}$Te and $^{98}$Zr). They form two
cascades of three $\gamma$-rays in mutual coincidences, which look like the new 
cascade of $^{125}$Sn (located on the top of the 27/2$^-$ state because of the coincidence 
relationships with its yrast transitions, see Fig.~\ref{schema125}). 
Thus we have tentatively placed the two new cascades above the long-lived 27/2$^-$ 
state of $^{123}$Sn (see Fig.~\ref{schema123}).

All the transitions observed  in $^{123}$Sn are given in 
Table~\ref{gammas123Sn}. In our experiments, a small number of events which contain 
the low-lying transitions of $^{123}$Sn are registered 
since the prompt fold of the corresponding cascades is strongly lowered   
by the long half-life of the 27/2$^-$ state (34~$\mu$s). Thus the intensities
of the transitions located below that state could not be given relative to that of
the ones located above the isomeric states. Nevertheless we have chosen to put the
$\gamma$-lines in Table~\ref{gammas123Sn}, as their energies are now known more 
precisely than previously~\cite{ma94}. 
The spin and parity of the new states will be 
discussed in Sec.~\ref{assign_spin_oddA}.  
\begin{table}[!h]
\caption{Properties of the $^{123}$Sn transitions. The excitation 
energy of the 23/2$^+$ state (written in bold) is
from Ref.~\cite{NNDC}.}
\label{gammas123Sn}
\begin{ruledtabular}
\begin{tabular}{rrccc}
E$_\gamma$(keV)\footnotemark[1]&I$_\gamma$\footnotemark[1]$^,$\footnotemark[2]&J$_i^\pi$$\rightarrow$J$_f^\pi$  &E$_i$(keV)&E$_f$(keV)\\
\hline
169.8(2)& \footnotemark[3]  &  27/2$^-$ $\rightarrow$ 23/2$^-$  & 2711.3 & 2541.5\\
261.4(3)  & 43(9)  & (35/2$^+$) $\rightarrow$ (31/2$^+$)           & 4378.0 & 4116.6 \\
279.1(2)& \footnotemark[3]  &  23/2$^-$ $\rightarrow$ 19/2$^-$  & 2541.5 & 2262.4 \\
823.6(3)  & 100  & (31/2$^+$) $\rightarrow$ (27/2$^+$)          & 4116.6 & 3293.0 \\
928.3(4)  & 3(1)  &~~~~~~  $\rightarrow$ (39/2$^+$)              & 7159.2 & 6231.0\\
1106.8(2)& \footnotemark[3] &  15/2$^-$ $\rightarrow$ 11/2$^-$  & 1106.8 & 0\\
1140.0(3) & 100  & (27/2$^+$) $\rightarrow$ 23/2$^+$             & 3293.0 & {\bf 2153.0}\\
1155.6(2)& \footnotemark[3] &  19/2$^-$ $\rightarrow$ 15/2$^-$   & 2262.4 & 1106.8 \\
1265.8(4) & 12(4)  &  ~~~~~~  $\rightarrow$ (35/2$^+$)             & 5643.8 & 4378.0\\
1515.3(5) & 6(2)  & 				              & 7159.2 & 5643.8 \\
1853.0(5) & 5(2)  & (39/2$^+$) $\rightarrow$ (35/2$^+$)           & 6231.0 & 4378.0 \\
&&&&\\
710.5(3)  & \footnotemark[4] & 				        & 5520.4& 4809.9 \\
735.0(3)  & \footnotemark[4] & 				        & 5477.6& 4742.6 \\
987.8(3)  & \footnotemark[4] &(35/2$^-$) $\rightarrow$ (31/2$^-$) &  4742.6& 3754.8 \\ 
991.6(3)  & \footnotemark[4] & (35/2$^-$) $\rightarrow$ (31/2$^-$)&  4809.9& 3818.3\\
1043.5(3) & \footnotemark[4] & (31/2$^-$)$\rightarrow$ 27/2$^-$   &  3754.8& 2711.3\\
1107.0(3) & \footnotemark[4] & (31/2$^-$) $\rightarrow$ 27/2$^-$  &  3818.3& 2711.3\\
\end{tabular}
\end{ruledtabular}
\footnotetext[1]{The number in parenthesis is the error in the last digit.}
\footnotetext[2]{The relative intensities are normalized to 
$I_\gamma(824) =100$.}
\footnotetext[3]{See text.}
\footnotetext[4]{Tentative attribution.}
\end{table}

\subsubsection{Angular momentum and parity values of the high-spin states of 
$^{119-125}$Sn}\label{assign_spin_oddA}

The statistics of our data related to the high-spin states of the odd-A Sn
nuclei is too low to perform $\gamma-\gamma$ angular correlation analyses.
Therefore, the spin assignments shown in 
Figs.~\ref{schema119}, \ref{schema125}, \ref{schema121}, and \ref{schema123} 
are based upon a few features:
\begin{itemize}
\item All the transitions with an energy $\gtrsim$ 1 MeV are assumed to have 
an $E2$ multipolarity.
\item The 223-keV transition of $^{121}$Sn is assumed to be $E2$ (see 
Sec.~\ref{$^{121}$Sn}). 
\item The 666- and 377-keV transitions of $^{121}$Sn, as well as the 824- 
and 261-keV transitions of $^{123}$Sn, are assumed to be $E2$, as the 
35/2$^+$ state with the $(\nu h_{11/2})^4(\nu d_{3/2})^1$ configuration is
expected to decay to the 23/2$^+$ state with the 
$(\nu h_{11/2})^2(\nu d_{3/2})^1$ configuration by
means of a cascade of three $E2$ transitions.
\end{itemize} 
Lastly, we can compute a few transition probabilities. 
As mentioned above, the half-life of the 2221-keV state of $^{121}$Sn is 
T$_{1/2} = 0.5(1) \mu$s, then the value of $B(E2;~23/2^+ \rightarrow 19/2^+)$ is
1.9(4)~$e^2fm^4$. 
The half-life of the 4378-keV state of $^{123}$Sn is 
T$_{1/2} <$ 30~ns, that leads to the value of the
reduced transition probability, 
$B(E2;~35/2^+ \rightarrow 31/2^+) > 15~e^2fm^4$.
 
\section{Discussion}\label{discuss}
\subsection{General features of $j^n$ configurations}\label{general}
The nuclear shell model (SM) describes the many-body nuclear
system in terms of a single-particle Hamiltonian representing the average
effect of the strong nucleon-nucleon interactions on a given nucleon, plus
residual interactions among a smaller number of particles, the valence nucleons 
near the Fermi surface, $H = H_0 + V_{res}$. 
The identification of states involving many identical nucleons in the 
same orbit $j$, i.e. states with the $j^n$ configuration, is a straightforward 
application of SM. 
These states are expected to exhibit typical features indicating properties
of the residual interaction. For instance it is now well known that, when it is 
assumed that only two-body forces contribute, $V_{res}(1,2)$, the interaction 
between $n$ nucleons in one shell $j$ can be expressed in terms of the 
two-particle matrix elements. Moreover, when using the seniority number, $v$ (which
can be defined as the number of unpaired nucleons\footnote{For a recent review 
on the use of the seniority quantum number in many-body systems, see 
Ref.~\cite{va10}.}), we can relate the two-body interaction matrix elements of 
seniority-$v$ states in the $j^n$ configuration to the matrix elements in the 
$j^v$ configuration. 

To illustrate these features, it is instructive to look at results of
calculations performed many years ago on the proton $(h_{11/2})^n$ 
configurations~\cite{la81}.   
The spectra associated with $n=$3, 4, 5,
and 6, the latter corresponding to mid-shell\footnote{Due to the
particle-hole symmetry, the spectrum associated to the $(\pi h_{11/2})^n$ 
configuration is the same as the $(\pi h_{11/2})^{12-n}$ one. Thus it is
sufficient to discuss the cases with $n \le 6$.}, were calculated
using the residual interactions taken from the experimental spectrum of 
$^{148}_{~66}$Dy$^{}_{82}$ (with $n=$2). For $n \ge$ 3, there are
often several states with the same angular momentum. If the two-body residual 
interactions conserve seniority, the latter can be used as a quantum number to
characterize each state. 
It is worth recalling that to be diagonal in the seniority scheme, a condition 
involving the five diagonal matrix elements
of the $(h_{11/2})^2$ interaction has to be fulfilled~\cite{ta93}. While this condition 
is not satisfied by most general two-body
interactions, it is fulfilled in the present case~\cite{la81}. Thus the 
predicted spectra of $(h_{11/2})^n$ configurations are very similar, whatever 
the number of nucleons. 

\begin{figure}[!h]
\includegraphics*[width=8.5cm]{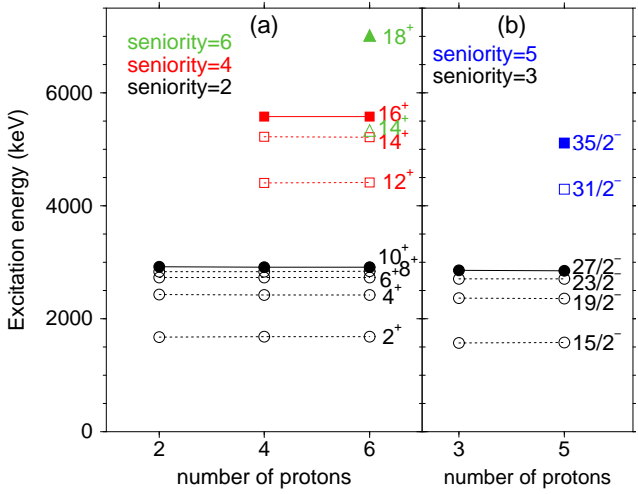}
\caption{(Color online) Evolution of the yrast states with the $(\pi h_{11/2})^n$ configuration as a
function of $n$, the number of protons occupying the orbit (see text). 
(a) Even number of protons. The states
with one broken pair ($v$=2) are drawn in black, with two broken pairs ($v$=4)
in red and three broken pairs ($v$=6) in green. (b) Odd number of protons. 
The states with one broken pair ($v$=3) are drawn in black, with two broken 
pairs ($v$=5) in blue.
}
\label{h11_seniority}      
\end{figure}

This is illustrated by the theoretical results obtained for the yrast states 
of $(\pi h_{11/2})^n$ configurations with even $n$, drawn in 
Fig.~\ref{h11_seniority}(a).
Above the first multiplet (I$^\pi$=2$^+$,.., 10$^+$) of 
$v=$2 seniority, we observe a second group of states (I$^\pi$=12$^+$, 14$^+$,
and 16$^+$) with $v$=4. 
When breaking a third $h_{11/2}$ pair, we obtain the 
highest-spin value available in this orbit, I$^\pi$=18$^+$. It is worth noting
that the $v$=6 14$^+$ state is located below the $v$=4 16$^+$ state, that has
a paramount importance for the decay of the 16$^+$ state. While the 
$16^+ \rightarrow 14^+_2$ transition is allowed, the one 
between states of the same seniority is forbidden as the orbit is half-filled,
i.e. $B(E2; 16^+ \rightarrow 14^+_1)$ =0. 
As a result, the value of the half-life of the 16$^+$ state of the $(\pi h_{11/2})^6$ 
configuration can be even lower than the one of  the 16$^+$ state  of the 
$(\pi h_{11/2})^4$.

The results obtained for the yrast states of $(\pi h_{11/2})^n$ configurations
with odd $n$ are drawn in Fig.~\ref{h11_seniority}(b). While the energy of
the 27/2$^-$ state is close to the one of the 10$^+$, the highest-spin states
obtained for an odd number of nucleons are located at lower energy than the
ones obtained for an even number. Thus we expect a more compressed spectrum when
the number of nucleons occupying the $j$ orbit is odd.

Unfortunately the experimental behaviors of the $(\pi h_{11/2})^{4,5,6}$ 
configurations have never been tested, the nuclei of interest lying very close 
to the proton drip-line.  Since they are only produced in reactions with very low 
cross sections, their high-spin states could not be identified.

\begin{figure}[!h]
\includegraphics*[width=8cm]{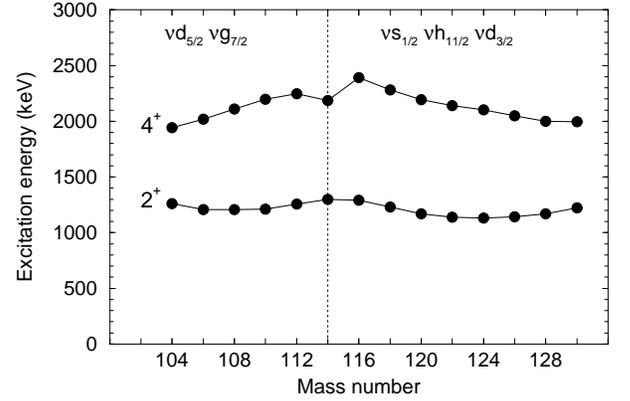}
\caption{Evolution of the 2$^+_1$ and 4$^+_1$ states of the Sn isotopes, as a
function of the mass number.
}
\label{2et4plus}      
\end{figure}

Pure $j^n$ configurations occur in a very few nuclei, since an 
orbit $j$ is rarely bounded by two gaps in energy. The closeness of several 
orbits leads to configuration mixings, nevertheless some of the features 
due to seniority are found to survive. For instance, there are in semi-magic 
nuclei fairly constant spacings between the 0$^+$ ground state and some 
$v=2$ states, even though large changes in configuration mixings
occur. The Sn nuclei provide a typical example, the energies of their first 
two states, (I$^\pi$=2$^+$ and 4$^+$) do not vary much across the
major shell ($54 < N < 80$), while the neutron orbits evolve from [$\nu
d_{5/2}, \nu g_{7/2}$] for $N < 64$ to 
[$\nu s_{1/2}, \nu h_{11/2}, \nu d_{3/2}$] for $N > 64$ (see Fig.~\ref{2et4plus}). 

The highest-spin states of the Sn
isotopes with $N > 64$ do contain a large $(\nu h_{11/2})^n$ component, since the
two low-$j$ orbits cannot afford large spin values. Therefore their study as a
function of the neutron number gives us the opportunity to explore the main
features of the  $(\nu h_{11/2})^{4,5,6}$ configurations. However such a work
is restricted to  $A \ge 120$ since the yrast states of the 
lightest-A Sn isotopes are dominated by a collective band coming from 2p-2h
excitations across the $Z=50$ gap, which hampers identifying states coming
from the $(\nu h_{11/2})^n$ configurations.
The maximum values of angular momentum obtained for various configurations 
with several broken pairs are given in Table~\ref{spinmax}. 
\begin{table}[!h]
\caption{Various configurations with
several broken pairs, expected in heavy-$A$ Sn isotopes.}
\label{spinmax}
\begin{ruledtabular}
\begin{tabular}{ccc}
configuration&$I^\pi_{max}$&nucleus\\
\hline
$(\nu h_{11/2})^2$	& 10$^+$    &even-$A$	\\
$(\nu h_{11/2})^3$	& 27/2$^-$  &odd-$A$ 	\\
$(\nu h_{11/2})^4$	& 16$^+$    &even-$A$	\\
$(\nu h_{11/2})^5$	& 35/2$^-$  &odd-$A$ 	\\
$(\nu h_{11/2})^6$	& 18$^+$    &even-$A$ 	\\
&&\\
$(\nu h_{11/2})^1(\nu d_{3/2})^1$	& 7$^-$		&even-$A$ \\
$(\nu h_{11/2})^2(\nu d_{3/2})^1$	& 23/2$^+$	&odd-$A$\\
$(\nu h_{11/2})^3(\nu d_{3/2})^1$	&15$^-$		&even-$A$\\
$(\nu h_{11/2})^4(\nu d_{3/2})^1$	& 35/2$^+$	&odd-$A$\\
$(\nu h_{11/2})^5(\nu d_{3/2})^1$	&19$^-$		&even-$A$\\
$(\nu h_{11/2})^6(\nu d_{3/2})^1$	&39/2$^+$	&odd-$A$\\
\end{tabular}
\end{ruledtabular}
\end{table}

\subsection{High-seniority states of the Sn isotopes}

\begin{figure}[!t]
\includegraphics*[angle=-90,width=8.5cm]{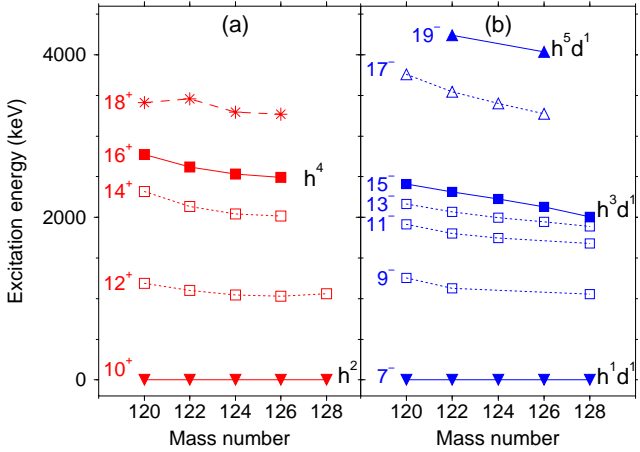}
\caption{(Color online) Evolution of the highest-spin states of the even-$A$ 
Sn isotopes, as a
function of the mass number (this work and Ref.~\cite{pi11} for $^{128}$Sn). The 
state having the maximum angular momentum of
each configuration is drawn with a filled symbol. (a) Excitation energies of the 
positive-parity states above the 10$^+$ level. For the peculiar behavior of the 
18$^+$ states (drawn with asterisks), see text. (b) 
Excitation energies of the negative-parity states above the 7$^-$ level. 
}
\label{seniorite_even}      
\end{figure}

The systematics of the excitation energies 
of the highest-spin states in the $^{120-128}$Sn even-$A$ isotopes are shown in 
Fig.~\ref{seniorite_even}. Noteworthy is the
fact that the highest-energy state of $^{122,126}$Sn measured in the present work
is assigned to be the 
19$^-$ state which is expected above the 17$^-$ state, knowing that the 
19$^-$-17$^-$ distance
in energy is most likely lower than that the 17$^-$- 15$^-$ one.

The evolution of the positive-parity states is very smooth. Likewise, the energies 
of the negative-parity states display a very regular behavior. 
This indicates that their main configurations are the same, whatever the number
of neutrons. Being very close in energy
for $69 \le N \le 81$, the three neutron orbits 
($\nu s_{1/2}, \nu h_{11/2}, \nu d_{3/2}$) are gradually filled together. 
Then the occupation number of the $\nu h_{11/2}$ subshell does not change by 
two units from one even-$A$ isotope to the next one and the maximum value of 
seniority does not display the stepwise behavior as a function of $A$, shown in
Fig.~\ref{h11_seniority}.

Taking into account the maximum values of angular momentum given in 
Table~\ref{spinmax}, it is tempting to assume that the positive-parity states have 
the $(\nu h_{11/2})^4$ and $(\nu h_{11/2})^6$ configurations, and the
negative-parity states, the $(\nu h_{11/2})^3(\nu d_{3/2})^1$ and 
$(\nu h_{11/2})^5(\nu d_{3/2})^1$ configurations.  
Nevertheless we can notice immediately the peculiar 
behavior of the 18$^+$ states: (i) the irregular variation of its excitation
energy as a function of $A$, and (ii) the low value of the 18$^+$-16$^+$ gap 
in energy as compared to the results given in Fig.~\ref{h11_seniority}. 
Thus the main configuration of the  18$^+$ states is likely not the $h^6$ 
configuration (this will be discussed in the next section).
All the assignments written in Fig.~\ref{seniorite_even} are corroborated by the
results of shell model calculations presented in Sec.~\ref{calculSM}.

As above mentioned, when using the proton $h_{11/2}^2$ 
residual interactions extracted from the $^{148}$Dy spectrum, the 14$^+$ state 
with seniority $v=$6 is located below the 16$^+$ state with seniority $v=$4. 
Then the 16$^+$ state decays predominantly towards 
that 14$^+_2$ state when the subshell is half-filled, since the decay towards 
the 14$^+_1$ state is hindered. As for the Sn isotopes, we did not observe any
delayed component in the $\gamma$-ray cascades located below the 16$^+$ states,
meaning that their half-lives are smaller than 30~ns. This would lead to 
$B(E2; 16^+ \rightarrow 14^+) > 0.7~e^2fm^4$ for $^{124}$Sn, for instance. This
limit is close to the low value of $B(E2; 10^+ \rightarrow 8^+)$ in
$^{122,124}$Sn, which is due to the half-filling of the $\nu h_{11/2}$ 
orbit~\cite{ma94} (see below). Thus in order to determine whether the 14$^+$ states
measured in the Sn isotopes have the seniority $v=$4 or $v=$6, we would need to know a
more precise limit of the 16$^+$ half-lives.

The $E2$ decays of the isomeric 15$^-$ states identified in the even-$A$  
$^{120-126}$Sn isotopes allow us to confirm their main configuration, 
$(\nu h_{11/2})^3(\nu d_{3/2})^1$.  
It is well known that the sign of the $E2$ transition amplitude between two states with the same
seniority  depends on the occupation rate of the orbit, being positive for low
values and negative for high values. Thus the behavior of 
$\sqrt{B(E2; 10^+ \rightarrow 8^+)}$ as a function of the Sn mass number was 
used to determine the half-filling of the $\nu h_{11/2}$ orbit, i.e. when 
particle and hole contributions
exactly cancel one another. That happens for $A=$123 or $N=$73~\cite{ma94}.
Results obtained in odd-$A$ Sn isotopes corroborated this value~\cite{br92,lo08}. Indeed the transition 
probability between two states with seniority $v=3$, such as   
$B(E2; 27/2^- \rightarrow 23/2^-)$, displays the same behavior with regard to 
the orbit filling as the one between two states with seniority $v=2$, such as 
$B(E2; 10^+ \rightarrow 8^+)$. Nevertheless in order to plot the $\sqrt{B(E2)}$ values of both
the even- and odd-A Sn isotopes in the same graph, one has to compensate 
the fractional parentage coefficients entering in the expression of  
the states with seniority $v=3$~\cite{ta93}: the $B(E2)$ values corresponding to
$v=3$ have to be multiplied by 0.264~\cite{ma89}. 

\begin{figure}[!h]
\includegraphics*[width=6cm]{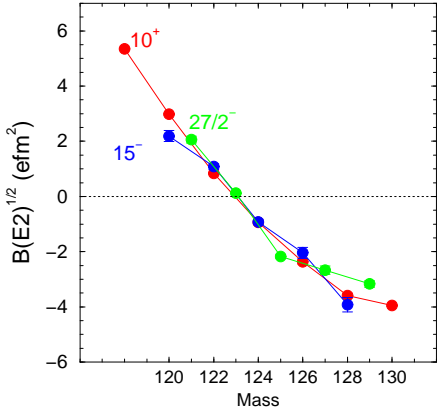}
\caption{(Color online) E2 transition amplitudes for the ($\nu h_{11/2})^{n}$
isomeric transitions in the Sn isotopes, $n=2$ for the decay of the 10$^+$
states, $n=3$ for the decays of the 27/2$^-$ and 15$^-$ states. 
The experimental values for the 27/2$^-$ states (in green) and for the 15$^-$ states 
(in blue) have been normalized (see text).
}
\label{racBE2}      
\end{figure}

The E2 transition amplitudes for these two sets of isomeric transitions are 
drawn in Fig.~\ref{racBE2}, as well as the results of the 
15$^-$ $\rightarrow$ 13$^-$ transitions obtained in the present work.
The latters follow exactly the same trend as those of the 
$10^+ \rightarrow 8^+$ transitions and the  $27/2^- \rightarrow 23/2^-$ ones
This confirms that the main configuration of the  15$^-$ and 13$^-$ states of 
the even-$A$  $^{120-126}$Sn isotopes is $(\nu h_{11/2})^3(\nu d_{3/2})^1$. 

\begin{figure}[!t]
\includegraphics*[angle=-90,width=7.6cm]{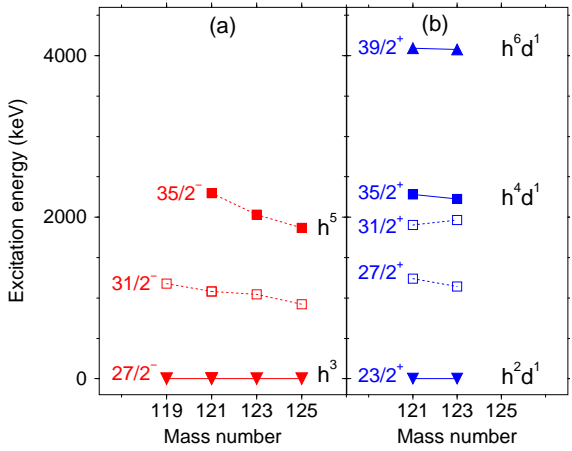}
\caption{(Color online) Evolution of the highest-spin states of the odd-$A$ Sn 
isotopes, as a function of the mass number (this work). The 
state having the maximum angular momentum of
each configuration is drawn with a filled symbol. 
(a) Excitation energies of the negative-parity states above 
the 27/2$^-$ level. (b) Excitation energies of the 
positive-parity states above the 23/2$^+$ level.
}
\label{seniorite_odd}      
\end{figure}

The excitation energies of the negative-parity states above the 27/2$^-$ level in
$^{119-125}$Sn are drawn in Fig.~\ref{seniorite_odd}(a). 
The highest-spin states of the seniority $v=$5 are more compressed in energy than
the ones of seniority $v=$4 shown in Fig.~\ref{seniorite_even}(a), as predicted for the 
$(\pi h_{11/2})^n$ configuration 
(see Sec.~\ref{general} and Fig.~\ref{h11_seniority}). 

The positive-parity states above the 23/2$^+$ level in $^{121}$Sn 
and $^{123}$Sn are due to the breaking of two and three neutron pairs in the $\nu
h_{11/2}$ orbit [see Fig.~\ref{seniorite_odd}(b)]. 
Their almost constant energies  
indicate once more that the three neutron orbits close to the
Fermi level are gradually filled, so the level energies do not depend very much on
the total number of neutrons.

\subsection{Results of shell model calculations}\label{calculSM}
We have performed shell-model calculations using the ANTOINE 
code~\cite{ca99}, the calculational details being the same as those described in 
Ref.~\cite{si09}. Since the present work mainly involves high-spin states, we have
restricted the calculations to the excited states with spin values higher than
7$^-$/10$^+$ for the even-$A$ isotopes and 23/2$^+$/27/2$^-$ for the odd-$A$.
In this section, we only present results of $^{125}$Sn and $^{126}$Sn, as those
of the lower-$A$ isotopes are nearly the same.

\begin{figure}[!h]
\includegraphics*[width=8.5cm]{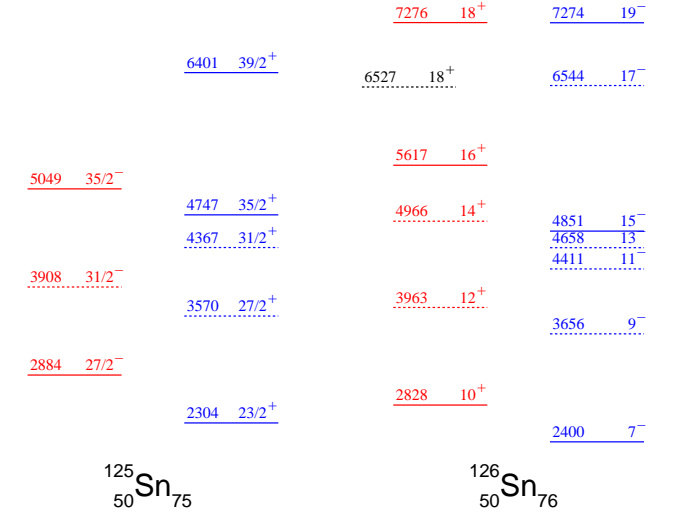}
\caption{(Color online) High-spin levels of $^{125}$Sn
and $^{126}$Sn predicted by the SM calculations (see text).
}
\label{schemaSM}      
\end{figure}

The yrast spectra above the long-lived isomeric states are shown in 
Fig.~\ref{schemaSM}.
The angular momenta of the states drawn with dotted lines comprise components 
involving a broken pair in the low-$j$ orbits, $\nu d_{3/2}$ or $\nu s_{1/2}$. 
The comparison of results given in Fig.~\ref{schemaSM} with those
obtained for a  pure $(h_{11/2})^n$ configuration (see Fig.~\ref{h11_seniority}) 
shows the influence of the low-$j$ neutron orbits in the energies of these
states. For instance, the 31/2$^-$ state is now located at mid-distance between the  
35/2$^-$ and the 27/2$^-$ states (see Fig.~\ref{schemaSM}). In the same manner, 
the 12$^+$ state is located at mid-distance between the 14$^+$ and the 10$^+$ states,
while it is located nearer the 14$^+$ state in the first calculation.

On the other hand, the states drawn with solid lines are due to the complete
alignment of the $h_{11/2}$ momenta of the neutron belonging to the broken pairs 
({\it cf.} the configurations given in Table ~\ref{spinmax}). For instance,  
the main configuration of the 35/2$^+$ state of $^{125}$Sn is 
$(\nu s_{1/2})^0 (\nu d_{3/2})^3 (\nu h_{11/2})^8$ (61\%), i.e. the breaking of
two $\nu h_{11/2}$ pairs, and the main configuration of
the 39/2$^+$ state is 
$(\nu s_{1/2})^2 (\nu d_{3/2})^3 (\nu h_{11/2})^6$ (79\%), i.e. the breaking of
three $\nu h_{11/2}$ pairs. Results obtained for $^{126}$Sn are very similar,
except for the 18$^+$ yrast state (drawn in black in Fig.~\ref{schemaSM}). 
Its main configuration is 
$(\nu s_{1/2})^1 (\nu d_{3/2})^3 (\nu h_{11/2})^8$ (59\%), the 
$(\nu s_{1/2})^2 (\nu d_{3/2})^4 (\nu h_{11/2})^6$ component being well weaker 
(11\%). Conversely, the main component of the 18$^+_2$ state (drawn in red
in Fig.~\ref{schemaSM}) is $(\nu s_{1/2})^2 (\nu d_{3/2})^4 (\nu h_{11/2})^6$
(79\%), i.e. the breaking of three $\nu h_{11/2}$ pairs. 
Thus in $^{126}$Sn, the breaking of the third $\nu h_{11/2}$ pair, which only 
provides an angular momentum of 2$\hbar$, competes 
unfavorably with the breaking of a low-$j$ pair, which can also provide an angular 
momentum of 2$\hbar$, either from the $(\nu s_{1/2})^1 (\nu d_{3/2})^3$
configuration or from the $(\nu d_{3/2})^2$ one.
It is worth noting that such a process does not occur in the 39/2$^+$ state of 
$^{125}$Sn, because of the blocking of the $\nu d_{3/2}$ orbit by the odd neutron.

These theoretical results are in good agreement with the experimental ones, 
particularly the
various distances in energy between successive states are well reproduced.
First, the 14$^+$-12$^+$ distance is found to be close to the 
12$^+$-10$^+$ one, as measured experimentally. Similar results are obtained for
the 35/2$^-$-31/2$^-$ and 31/2$^-$-27/2$^-$ spacings which are well predicted. 
Secondly, the bunching of the three states
with $I^\pi=$11$^-$, 13$^-$, and 15$^-$  is well reproduced, as well as the large
gap between the 15$^-$ and 17$^-$ states.
Lastly, the peculiar behavior of the experimental 18$^+$ states is likely due 
to the breaking of a low-$j$ neutron pair, which is
expected to evolve as a function of $A$ since the $\nu s_{1/2}$ orbit gets farther
away from the neutron Fermi level as $N$ is increasing.

In summary, we have identified states having the maximum value of angular momentum of either the 
$(\nu h_{11/2})^n$ configurations or the $(\nu h_{11/2})^n (\nu d_{3/2})^1$ ones, 
with $n$=1, 2, 3, 4, 5, and 6, in the even-$A$ and 
odd-$A$ Sn isotopes (see Table~\ref{spinmax} and also Figs.~\ref{seniorite_even} 
and \ref{seniorite_odd}).  The only missing state is the 18$^+$ state from the 
$(\nu h_{11/2})^6$ configuration, which is not yrast.

\subsection{Search for high-seniority states from other $j^n$ configurations}\label{autrescas}
While there are numerous examples of states with seniority $v=$2 in the literature,
states with seniority $v=$4 are scarce. The 10$^+$ and 12$^+$ states of the   
$(\pi g_{9/2})^4$ configuration have been identified in the two $N=$50 isotones, 
$^{94}$Ru and $^{96}$Pd~\cite{NNDC}. On the other hand, the highest-spin level 
measured in the $(\nu g_{9/2})^4$ configuration ($^{72,74}$Ni$_{44,46}$) is the 
8$^+$ state 
of seniority $v=$2. As for the $(\pi h_{11/2})^n$ configuration, which is expected
in the $N=$82 isotones having $Z \ge 68$, only one experimental result has been 
obtained as the nuclei of interest are close to the proton drip line: The  $v=$4 
16$^+$ state of the $(\pi h_{11/2})^4$ configuration is identified in 
$^{150}_{~68}$Er~\cite{NNDC}.

The next high-$j$ orbital is $i_{13/2}$, which is located in the middle of the 
82-126 major shell. The excitation energy of the 13/2$^+$ state measured in
$^{199,201}$Pb is around 500 keV above the 5/2$^-$ and 3/2$^-$ states, meaning 
that the main configurations of the high-spin states expected in the Pb isotopes 
are likely complex. Indeed several $v=$4 states which were measured in the
$^{200,202,204}$Pb nuclei~\cite{NNDC}, have the configurations 
$(\nu i_{13/2})^2(\nu f_{5/2})^2$ (16$^+$) and $(\nu i_{13/2})^3(\nu f_{5/2})^1$ 
(17$^-$, 19$^-$), but the breaking of two or three pairs of $i_{13/2}$ neutrons has
never been observed. We have to recall that the high-spin level schemes of these 
Pb nuclei display coexisting collective structures which make difficult to identify
the states having $(\nu i_{13/2})^n$ configuration, with $n > 3$. 

In the Sn nuclei, several high-spin states having a configuration $j^v$, with $v=$5 
and 6, are identified for the first time in the present work. Such an observation
is unique in the whole nuclear chart, because several prerequisites are needed
which seem to be only fulfilled by the heavy-$A$ Sn 
nuclei and the $\nu h_{11/2}$ orbit: (i) the nuclei have to be spherical, without 
shape coexistence leading to a vast number of high-spin states forming the 
yrast sequence, (ii) the spherical high-$j$ orbit has to be isolated from the 
others in order to be the main active one, (iii) the high-spin states of the 
nuclei of interest have to be populated with enough intensity to be measured.

\section{Summary and Conclusion}\label{summary}
Thanks to the high efficiency of the Euroball array, many high-spin states have
been identified in $^{119-126}$Sn isotopes. These nuclei have been produced as
fission fragments in two reactions induced by heavy ions: 
$^{12}$C+$^{238}$U at 90 MeV 
bombarding energy, $^{18}$O+$^{208}$Pb at 85 MeV. The level schemes have been
built up to 5-7~MeV excitation energy by analyzing triple $\gamma$-ray coincidence
data. Spin and parity values have been assigned to most of the high-spin states
using $\gamma-\gamma$ angular correlation results. In addition,  
the use of the fission-fragment detector, SAPhIR, has allowed us to identify new
isomeric states in the even-$A$ $^{120-126}$Sn isotopes.
All the observed states can be described in terms of several broken neutron pairs  
occupying the $\nu h_{11/2}$ orbit. The maximum value of angular momentum 
available in this high-j shell, i.e. for its mid-occupation and the breaking of 
the three pairs (seniority $v$=6), has been identified. 

This process, which is observed for the first time in 
spherical nuclei, has some similarities with the one involved in some 
deformed nuclei.
Several $N=90$ isotones show evolution of collectivity with spin. While the 
nuclei are prolate for $I$=0, the lowest energy states at $I$=40-50 are due to
oblate shape. There the nuclei can be considered as a closed core of $^{146}$Gd
and several additional valence particles in the $j$ orbits lying above the
$Z=$64 and $N=$82 closures. Then the  observed band structure terminates at the
maximum value of spin available to the valence nucleons. For instance, in    
$^{158}_{~68}$Er, band termination occurs at spin 46$\hbar$~\cite{tj85}, when all the 12 
valence nucleon spins are aligned, 
$(\pi h_{11/2})^4 (\nu f_{7/2})^3 (\nu h_{9/2})^3 (\nu i_{13/2})^2$, i.e. a
configuration with 6 broken pairs, but involving 4 different orbits. 
\begin{acknowledgments}
The Euroball project was a collaboration among France, 
the United Kingdom, Germany,
Italy, Denmark and Sweden. The first experiment has been performed under 
U.E. contract (ERB FHGECT 980 110) at Legnaro. The second one has 
been supported in part by the EU under contract HPRI-CT-1999-00078 (EUROVIV). 
We thank many colleagues for their
active participation in the experiments, Drs. A.~Bogachev, A.~Buta, J.L.~Durell, 
Th.~Ethvignot, F.~Khalfalla, I.~ Piqueras, A.A.~Roach, A.G.~Smith and B.J.~Varley.
We thank the crews of the tandem 
of Legnaro and of the Vivitron, as well as M.-A.~Saettle
for preparing the Pb target, P.~Bednarczyk, J.~Devin, 
J.-M.~Gallone, P.~M\'edina and D.~Vintache
for their technical help during the second experiment. 
We are indebted to K. Sieja and F. Nowacki for providing us with their effective 
nucleon-nucleon interactions.

\end{acknowledgments}

\end{document}